\documentclass[%
 reprint,
 amsmath,amssymb,
 aps,
pra,
floatfix,
]{revtex4-2}

\usepackage{graphicx}
\usepackage{dcolumn}
\usepackage{bm}

\usepackage{amsfonts}
\usepackage{amssymb}
\usepackage{physics}
\usepackage{geometry}
\usepackage{fancyhdr}
\usepackage{graphicx}
\usepackage{subcaption}
\usepackage{ragged2e}
\usepackage{tikz}
\usepackage{hyperref}
\usepackage[bb=libus]{mathalpha}

\begin{document}

\newcommand{\mat}[1]{\stackrel{\leftrightarrow}{#1}}
\newcommand{\comega}{\tilde{\omega}}
\newcommand{\cOmega}{\tilde{\Omega}}

\preprint{APS/123-QED}

\title{Multimode Entangled Squeezed Light Generation and Propagation\\ in a Coupled-Cavity Photonic Crystal}

\author{Dylan van Eeden}
 \email{dylan.vaneeden@queensu.ca}
\author{Marc M. Dignam}%

\affiliation{
    Department of Physics, Engineering Physics, and Astronomy\\
    Queen's University, Kingston, Ontario K7L 3N6, Canada
}%
\date{\today}

\begin{abstract}
Entangled multi-mode squeezed states of light have a wide variety of applications in quantum information systems, particularly in the generation of non-Gaussian states of light, which are central to continuous-variable quantum computing. Although theoretical approaches exist to model the nonlinear generation of one- and two-mode entangled states of light in ring resonator systems, these approaches are difficult to implement in modeling more complicated many-cavity systems. In this work, we present an efficient and accurate theoretical approach to modeling the generation and propagation of quantum states of light in lossy coupled-cavity systems containing a two- or three-mode nonlinear resonant structure. Our approach is general and computationally viable even in systems with hundreds of modes. We apply our method to the design and modeling of a multimode photonic crystal coupled-cavity system for the generation of entangled squeezed states of light on-chip. The system consists of a three-mode resonant structure coupled to three coupled-resonator optical waveguides (CROWs) in a square lattice silicon photonic crystal slab. The computational speed of the method allows us to efficiently optimize the system such that the signal and idler light in the two output CROWs remains entangled even after propagating tens of cavities down the CROWs. 
\end{abstract}

\maketitle


\let\clearpage\relax
\section{Introduction}\label{sec:I}
Entangled squeezed thermal states of light \cite{Pfister2020,Adesso2014} are sought for their entanglement and noise properties. They have applications in continuous-variable quantum computing \cite{Alexander2024,Feldman2024,Menicucci2006}, quantum key distribution \cite{Leverrier2011}, quantum-enhanced sensing \cite{Dorfman2016, LIGO, Schnabel2017}, and the generation of non-Gaussian entangled states of light \cite{Tomoda2024}. Consequently, there is considerable interest in modeling, designing, and optimizing nanophotonic platforms for the generation of entangled quantum states of light.

\begin{figure*}[!ht]
    \centering
    \includegraphics[width=\textwidth]{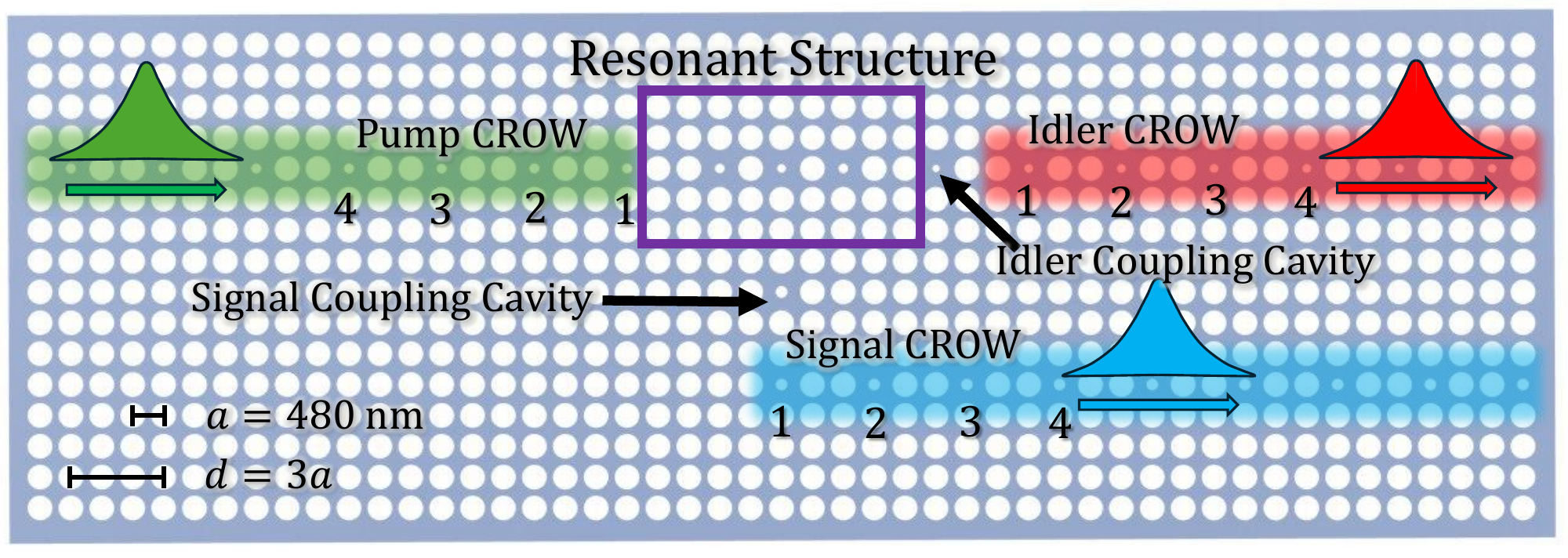}
    \caption{\justifying Schematic of the designed photonic crystal coupled-cavity system in the vicinity of the resonant structure (outlined by the purple box) with key features highlighted. White regions indicate air holes etched into the silicon slab (pale blue). A pump pulse is injected via the pump CROW (green, left of RS), and the generated light leaves the RS via the signal (blue, lower-right of RS) and idler (red, right of RS) CROWs in spatially separated pulses. The numbers below the three CROWS, indicate the cavity labeling convention that will be used in Sec. \ref{sec:IV}. Note that the three CROWs used in the simulation are much longer than shown here, with each containing over 50 coupled cavities.}
    \label{fig:PCCCS}
\end{figure*}

Entangled, multi-mode, squeezed thermal states of light may be created by generating single-mode squeezed thermal states of light in an optical cavity \cite{Seifoory2017,Vendromin2020} and interfering them using a beam-splitter array to build multi-mode entanglement \cite{Madsen2022}. Alternatively, they can be generated directly via a nonlinear process described by a Hamiltonian containing terms with an operator of the form $(a^\dagger_p a^\dagger_q)^n,\ n\in\mathbb{N}^{>0}$, where $a_p^\dagger$ is the creation operator for photons in the $p^{th}$ mode. 

Nonlinear devices using microring resonators  have been explored both experimentally \cite{Silverstone2015, Tan2019, Zhang2021, Madsen2022} and theoretically \cite{Quesada2021, Mataji-Kojouri2024, Vendromin2020, Vendromin2021, Vendromin2022, Vendromin2023, Vendromin2024} for the generation of entangled squeezed states of light. Microring resonator systems can be fabricated with very high quality factors (Q) \cite{Zhang2021} and are relatively simple to design, fabricate, and couple to channel waveguides. Moreover, due to the relative simplicity of the ring modes as well as their coupling to the channel waveguide, accurate modeling of state generation in these systems can be achieved using only a few ring modes along with a well-established input-output formalism \cite{Zhang2021,Dignam2025}. There are, however, two important drawbacks of ring resonator systems.  First, they are relatively large, with typical ring radii of 50 to 100 microns.  Second, the rings support a large number of nearly equally spaced modes, which can lead to parasitic nonlinear processes that degrade the properties of the desired quantum state \cite{ Zhang2021, Seifoory2022} unless more complicated coupled-ring systems are employed \cite{Zhang2021} \cite{Larsen2025}. It is therefore salient to explore alternative platforms for multimode squeezed state generation, aiming to improve the footprint and optical properties.

One alternative to ring resonator systems 
are photonic crystal (PC) coupled cavity systems (CCSs).  Such systems have been employed for the generation of photon pairs \cite{Azzini2013} and have been investigated theoretically \cite{KamandarDezfouli2017,Seifoory2018,Seifoory2019} for the generation of squeezed states of light. Although the quality factors of the resonant modes in PC CCSs are generally an order of magnitude or more smaller than that of ring resonator systems, the resonant cavities (and mode volumes) are much smaller, with cavity sizes on the order of hundreds of nanometers. Moreover, they can be designed such that they only support the modes required for the desired nonlinear optical processes. Finally, employing coupled resonator optical waveguides (CROWs) for the output ports, signal filtering is integrated in the design, removing the need to include additional filters, which can add considerable loss \cite{Larsen2025}. 


In this work, we examine PC CCSs as an alternative platform to microring resonators for the generation of squeezed states (see Fig. \ref{fig:PCCCS}). 
The system consists of a three-mode, three-defect resonant structure (RS) coupled to three CROWs \cite{Yariv1999} in a square lattice silicon photonic crystal slab (see Fig. \ref{fig:PCCCS}). 
The key element of the system is the RS: it is comprised of three single-defect cavities and supports three resonant modes, one each at the signal, pump, and idler frequencies. The RS enhances the nonlinear spontaneous four wave mixing (SFWM) process that generates the desired entangled state. An entangled two-mode squeezed thermal state (2MSTS) is generated in the RS at the signal and idler frequencies via SFWM when excited by a Gaussian pump pulse with a frequency that is mid-way between the signal and idler frequencies. The generated light propagates out of the RS as two entangled pulses at the signal and idler frequencies in spatially-distinct CROWs that are designed to block any pump light. 


Any model of squeezed state generation in nanophotonic systems must include: 1) the nonlinear generation process; 2) the effects of loss; and 3) the coupling of the generated light into the output channel waveguide(s). In ring resonators, because the mode structure of the combined ring and channel waveguide system is relatively simple, with the loss being essentially the same for all modes, the nonlinear evolution and mode propagation can be treated accurately using an input-output formalism, with the loss accounted for either using phantom channels \cite{Quesada2021}, or using a density matrix formalism for the ring modes \cite{Dignam2025}.  However, in PC coupled cavity systems, the mode structure can be complicated, with the loss being very different in different modes, and with convoluted pump and signal propagation dynamics. Although one could, in principle, model such systems using a multi-mode density matrix formalism, such an approach quickly becomes computationally intractable because the number of Fock basis states required for convergence grows as $(N+1)^M$, where $M$ is the number of modes and $N$ is the maximum number of photons that may be generated in each mode. For example, in this work we study the generation and propagation of a squeezed thermal state in a system of 184 coupled cavities; modeling this state of light in such a system would require a minimum of $M=33$ modes \footnote{We arrive at the minimum mode number of $M=33$ by projecting the state generated in the RS onto the full-system QMs, and counting the number of full-system QMs with at least $1\%$ of the total number of photons.}. If we permit each of these 33 modes to have up to $N=2$ photons, then we require $3^{33}$, or more than $5\times10^{15}$ total Fock basis states. Such a problem is clearly intractable when working in a Fock basis.  


In this work, we develop a new theoretical approach to simulate the nonlinear generation of entangled multi-mode squeezed thermal states of light in systems of coupled lossy nanophotonic resonators in which there is a localized region -- the RS -- where the nonlinear interaction occurs. 
Our approach enables the simulation and study of quantum states of light in a complex PC CCS that cannot easily be modeled using traditional Fock-state based techniques.

\begin{figure}[h!]
    \centering
    \begin{tikzpicture}[grey_tb/.style={rectangle,draw=black!100,fill=black!20,thick,inner sep=5pt,minimum size=7mm}]
        \node [grey_tb] (1) at (0,1.5) {1: Compute Full-System and Subsystem QMs};
        \node [grey_tb] (2) at (0,0.5) {2: Classical Evolution of Pump Pulse};
        \node [grey_tb] (3) at (0,-0.5) {3: Nonlinear Generation of 2MSTS in RS};
        \node [grey_tb] (4) at (0,-1.5) {4: Quantum Evolution of Observables};
        \draw [color=black!100, line width=2pt, ->] (1) -- (2);
        \draw [color=black!100, line width=2pt, ->] (2) -- (3);
        \draw [color=black!100, line width=2pt, ->] (3) -- (4);
    \end{tikzpicture}
    \caption{\justifying Procedure flowchart for the four-step analysis used to calculate the quantum state of light in the output CROWS.}
    \label{fig:flowchart}
\end{figure}

We apply our approach to the simulation and optimization of the PC CCS shown in Fig. \ref{fig:PCCCS}, but note that our method can be applied to many other nanophotonic resonator systems. The simulation is carried out in four distinct steps, as shown in Fig. \ref{fig:flowchart}. First, we determine the lossy quasi-modes (QMs) of the full system under consideration. To accomplish this, we use finite difference time domain (FDTD) simulations to compute the electric field profile and quality factor of the QMs of the single-defect cavities from which we construct the PC CCS \cite{Fussell2007}. We then use a tight-binding approach to construct the QMs of the full system from these single-defect modes \cite{Fussell2007,Fussell2008,Dignam2012,KamandarDezfouli2014}. Second, we use this basis of system QMs to compute the classical evolution of the pump light (for a given initial pump field) and determine the pumping-field amplitude inside the RS as a function of time. Third, we compute the time evolution of density operator of the quantum state of light generated via nonlinear interaction in the RS. Fourth, we determine the quantum evolution of the generated state throughout the system by solving the Adjoint master equation after the pump has largely left the RS.  

 Our simulations show that strong squeezing can be obtained in the RS and that entanglement between the light and the signal CROWs persists for many tens of cavities down the waveguides. However, we find that entanglement between the signal and idler is significantly degraded in the CROWs. We analyze the source of this degradation and point to possible solutions.


The paper is organized as follows: in Section \ref{sec:II}, we provide a detailed description of the designed device and outline the design goals; in Section \ref{sec:III}, we introduce the four-step analysis procedure we developed to simulate the designed system; in Section \ref{sec:IV}, we present and discuss the results of the analysis; and finally in Section \ref{sec:con}, we conclude. 
\section{System Overview and Design Goals}\label{sec:II}

In this section, we describe the PC CCS that we have designed and optimized using our theoretical approach. 
We present our design process, outline the objectives of our design, and present the specifications of the designed PC CCS within which we will simulate the generation and propagation of multimode squeezed light. The details of the calculation of the coupled cavity modes are presented in Appendix \ref{AppA}.

The PC CCS consists of a three-mode RS  and three CROWs in a PC slab, shown schematically in Fig. \ref{fig:PCCCS}. The underlying PC slab is a 384 nm thick silicon (n=3.47) slab with air holes of radius 192 nm etched in a rectangular lattice with a lattice constant $a=480$ nm. This particular square lattice geometry of the PC was chosen because the single-defect cavities formed in this PC support a single relatively high-Q mode in the photonic band-gap of the PC \cite{Fussell2007}. 

We have designed the PC CCS to generate and propagate an entangled 2M STS with high brightness and robust entanglement. We have four main objectives for the design of the PC CCS. First, we need a RS with spatially-overlapping modes at the frequencies of the pump ($\omega_P$), signal ($\omega_S$), and idler ($\omega_I$), where $2\omega_p =\omega_s + \omega_I$ to enable SFWM between the pump, signal, and idler modes. Second, we need to be able to inject the pump light into the RS via a CROW such that there is a large pump intensity buildup inside of the RS.  Third, we require that the quantum state of light generated in the RS can couple out of the RS into two separate CROWs, with central frequencies $\omega_s$ and $\omega_I$ that do not allow propagation of the pump light. The key properties of the system are summarized in Table \ref{tab:systemProperties}.

\begin{figure}[!h]
    \centering
    \begin{subfigure}[r]{0.5\textwidth}
        \includegraphics[scale=0.45]{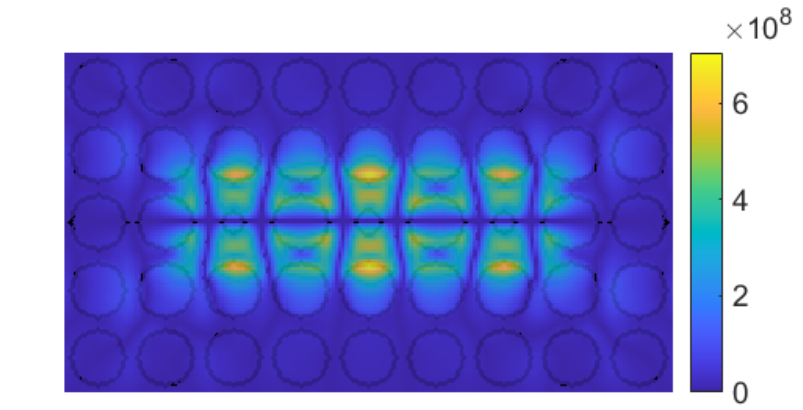}
        \caption{Idler, $\comega_I = 0.30610- i1.19\times10^{-5}\ (2\pi c/a)$}
    \end{subfigure}
    \begin{subfigure}[r]{0.5\textwidth}
        \includegraphics[scale=0.45]{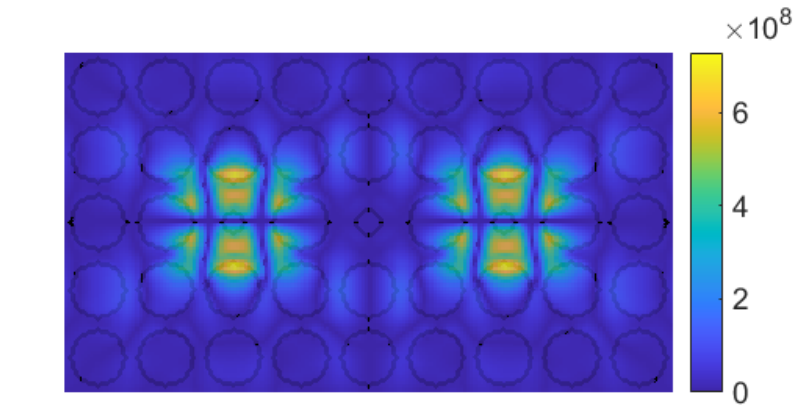}
        \caption{Pump, $\comega_P = 0.30763- i8.17\times10^{-6}\ (2\pi c/a)$}
    \end{subfigure}
    \begin{subfigure}[r]{0.5\textwidth}
        \includegraphics[scale=0.45]{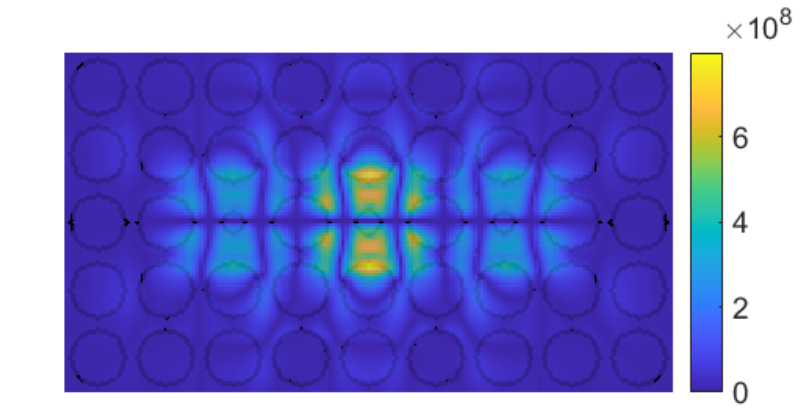}
        \caption{Signal, $\comega_S = 0.30916- i3.69\times10^{-6}\ (2\pi c/a)$}
    \end{subfigure}
    \caption{\justifying Plot of $|C_{j,x}(x,y,0)|^2$ for the three RS QMs. The area of each subfigure is enclosed by the purple rectangle in Fig. \ref{fig:PCCCS}. The complex frequency for each of the QMs (when the CROWs are absent) are given below each plot.}
    \label{fig:RSQM}
\end{figure}

The PC CCS was designed starting with the RS, outlined in solid purple in Fig. \ref{fig:PCCCS}. The RS contains three cavities: the outer two have air holes with radii of 74 nm and the center has an air hole with a radius of 80 nm. This three-cavity structure supports three resonant QMs, denoted by $\textbf{C}_j(\textbf{r})$, which in increasing order of resonance frequency are the idler, pump, and signal modes. The hole sizes were chosen by scanning all configurations in which the wing cavities were identical (see Fig. \ref{fig:freq_vs_r} in Appendix \ref{AppA}) until the resonance condition $2\omega_p =\omega_s + \omega_I$ was met.  In Fig. \ref{fig:RSQM}, we plot $|C_{j,x}(x,y,0)|^2$ for each of the three RS QMs and list their complex frequencies.  The operating frequencies for SFWM are given $\omega_j\equiv\Re{\comega_j}$, while the intrinsic power loss rate is given by $\Gamma'_j\equiv-2\Im{\comega_j}$ (see Table \ref{tab:systemProperties}). We note that the three RS QMs strongly overlap in space, which enables a strong SFWM interaction. 

\begin{figure}[htp]
    \centering
    \includegraphics[scale=0.35]{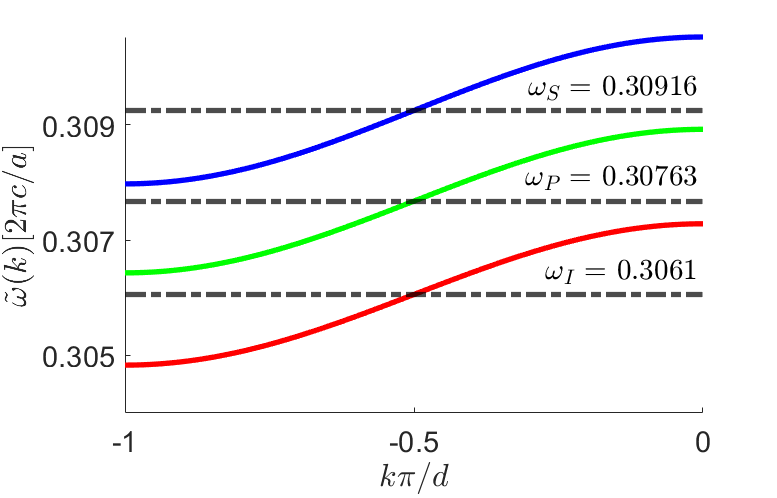}
    \caption{\justifying Dispersion for the CROW bands as calculated in the nearest-neighbor tight-binding approximation (see Appendix \ref{AppA}). Idler, pump, and signal CROW bands are shown in red, green, and blue, respectively, with corresponding RS QM frequencies displayed as dashed lines intersecting their corresponding CROW band.}
    \label{fig:crowbands}
\end{figure}
We next designed the three CROW structures used to couple the pump light into and generated light out of the RS. The dispersion curves for the three CROWs are shown in Fig. \ref{fig:crowbands}. The CROWS were designed so that the center frequency of each band matches the corresponding RS QM frequency (see Table \ref{tab:systemProperties}) and their bandwidths encompass only one of the pump, signal, or idler frequencies. This was achieved by constructing each CROW from a set of single-defect cavities with a resonant frequency equal to the frequency of the corresponding mode of the RS, and by setting the period to $d=3a$ (for all three CROWs). 
The resultant single-defect cavities chosen have radii of 66, 76, and 89 nm respectively for the idler, pump, and signal CROWs. Because the signal and idler CROWs can't pass light at the pump frequency, there will be no pump contamination in the output CROWS.  Moreover, because the pass bands of signal and idler CROWs do not overlap, the light exiting in the signal and idler CROWs will have no frequency overlap.  


\begin{table}[h!]
    \centering
    \begin{tabular}{l|c|c|c}
         Parameter &  Pump & Signal& Idler \\\hline\hline
         RS QM $\lambda$ (nm) & 1560 & 1553 & 1568\\
         RS QM $f$ (THz) & 192.1 & 193.1 & 191.2 \\
         RS QM $\omega\ (2\pi c/a)$ & 0.3076 & 0.3092 & 0.3061 \\
         RS QM intrinsic Q &18800 &41900 &12900 \\
         CROW Cav. Rad. (nm) & 76 & 89 & 66 \\
         CROW Ctr. Freq. (THz) & 192.3 & 193.2 & 191.2\\
        Coupling Cav. Rad. (nm) & -  & 91 & 55       
    \end{tabular}
    \caption{\justifying Summary of RS QM properties, defect cavity radii, and CROW properties for the PC CCS shown in Fig. \ref{fig:PCCCS}.}
    \label{tab:systemProperties}
\end{table}

\section{Theoretical Approach }\label{sec:III}

In this section, we present our theoretical approach to simulating the generation of quantum states of light via nonlinear processes in a PC CCS. The subsections describe the following four steps that we employ to simulate the generation and time evolution of the quantum state of light (see Fig. \ref{fig:flowchart}): A) Calculation of the quasimode basis states; B) Calculation of the classical pump evolution; C) Modeling the nonlinear generation of the quantum state in the RS; and D) Calculation of the free evolution of the quantum state of light in the full PC CCS. 
This section provides an overview of our approach; the details can be found in the appendices cited in the text. 

\subsection{Quasi-Mode Bases}
\begin{figure}[h]
    \centering
    \includegraphics[width=0.48\textwidth]{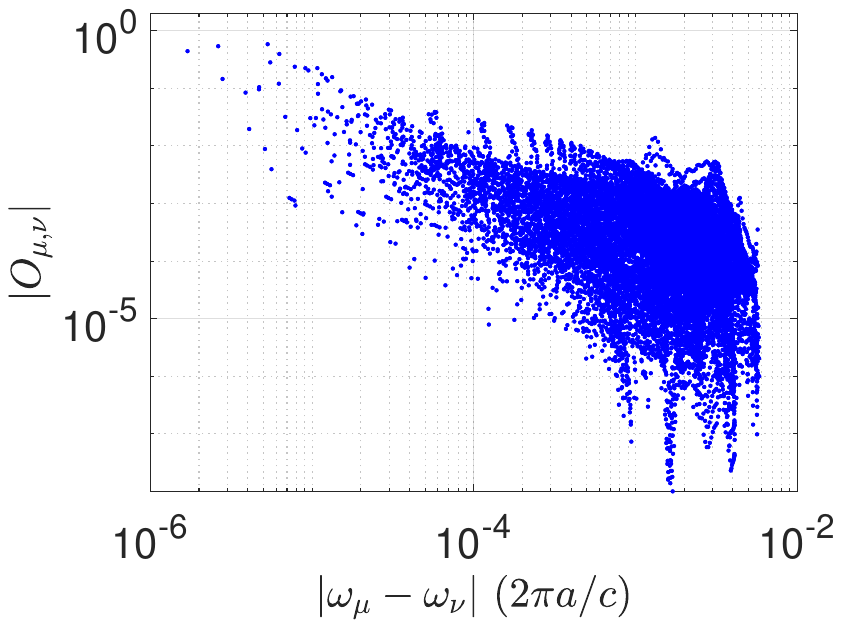}
    \caption{\justifying Absolute values of the QM overlaps $O_{\mu ,\nu}$ for the full system as a function of the absolute value of frequency difference between the $\mu$ and $\nu$ QMs.}
    \label{fig:FullSystemOverlaps}
\end{figure}

We begin by calculating the four different bases that are used in the simulation: the single-defect QMs $\textbf{M}_q(r)$, the Bloch QMs of the pump CROW $\textbf{F}_k(\textbf{r})$, the RS QMs $\textbf{C}_j(\textbf{r})$, and the full-system QMs $\textbf{N}_\mu(\textbf{r})$. We start by calculating the QMs (and complex frequencies) of the \textit{individual cavities} using FDTD. We then use these QMs in a tight-binding approach \cite{Fussell2007} to calculate the QMs of the different subsystems and of the entire system. For example, for the full system, the expansion takes the form
\begin{align}
    \label{eq:NmodeExp}
    \textbf{N}_\mu(\textbf{r})=\sum_q v_{q\mu} \textbf{M}_q(\textbf{r}),
\end{align}
where $v_{q\mu}$ are the elements of the invertible square matrix $\mat{V}$, which are determined (along with the QM complex frequencies $\comega_\mu$) by solving a generalized eigenvalue equation (see Appendix \ref{AppA} for details).
We note that this tight-binding approach makes it possible to find the QMs of very large coupled-cavity structures that would be extremely difficult and computationally expensive to determine using FDTD or finite element methods. In general, the QMs of a lossy coupled-cavity structure are non-orthogonal, as the problem is non-Hermitian. Although the QMs of the CROWS are orthogonal and the QMs of the RS are effectively orthogonal, the full-system QMs can be highly non-orthogonal: Fig. \ref{fig:FullSystemOverlaps} displays the overlap between the full-system QMs. This non-orthogonality affects the dynamics of the light in the system, as we shall discuss below. Mathematically, the effect of the non-orthogonality is to mix the QMs during the free evolution of the light.

\subsection{Pump Evolution}
Having determined all of the relevant QMs, we next calculate the classical evolution of the pump pulse as it propagates along the pump CROW and couples into the RS. At the initial time, $t=0$, we take the pump to be in a Gaussian superposition of Bloch modes $\textbf{F}_k(\textbf{r})$ in the pump CROW, well away from the RS. 
The positive frequency part of the electric field of the pump light at $t=0$ is given by
\begin{align}
    \label{eq:Epexp}
    \textbf{E}_P^{(+)}(\textbf{r},t=0)=i\sum_k\alpha_P^{(0)}(k)\sqrt{\frac{\hbar\tilde{\omega}_k}{2\epsilon_0}}\textbf{F}_k(\textbf{r}),
\end{align}
where the $\alpha_P^{(0)}(k)$ are the expansion coefficients, which take the Gaussian form
\begin{align}
    \label{eq:alpha_P}
    \alpha^{(0)}_P(k)=\sqrt{\frac{\kappa d n_p}{\sqrt{2\pi}}} e^{-(k-k_0)^2/\kappa^2}e^{-iq_0 kd},
\end{align}
where $n_p$ is the total number of photons initially in the pump pulse, $q_0$ is the index of the cavity at which the pulse is centered at $t=0$, and $\kappa$ is the $k$-space width parameter of the pulse. 
We then project this state onto the basis of the QM of the full system, which allows us to determine the time evolution of the pump in the full structure. We finally project the time-evolved pump field onto the basis of the three RS QMs to compute the time-dependent coherent state amplitude, $\alpha_{P}(t)$, of the pump in the RS pump mode, which is used in the describe in the following section. The full mathematical details of this procedure are given in Appendix \ref{AppB}. 

\subsection{Nonlinear Generation}
In this subsection, we present our approach to calculating the density operator of the two-mode squeezed thermal state generated in the RS of our system via SFWM. 
We make the undepleted pump approximation, \textit{i.e.}, we assume that the pump is classical and is unaffected by the nonlinear interaction.  This is a good approximation as long as the number of photons generated is much less than the number of photons in the pump in the RS, which, as we shall show, is the case in our simulations.

Restricting ourselves to the three modes of the RS, we expand the ladder operators, $\{d_j,\ d_j^\dagger\}$, for the $j^{th}$ RS QM ${\textbf{C}}_j(\textbf{r})$ as 
\begin{align}
    \label{eq:RSQMLOPS}
    d_j = \sum_{q=1}^3 u_{qj}a_q,\quad d_j^\dagger = \sum_{q=1}^3 u^*_{qj}a_q^\dagger,
\end{align}
where $\qty{a_q,\ a^\dagger_q}$ are the ladder operators for the $q^{th}$ single-defect cavity mode ${\textbf{M}}_q(\textbf{r})$, and the $u_{qj}$ are components of the matrix $\mat{U}$, which gives the mode expansion coefficients for ${\textbf{C}}_j(\textbf{r})$ (see Eq. (\ref{eq:Cexpansion})). In this work, we only consider the SFWM nonlinear interaction in which two pump photons are exchanged for one signal and one idler photon. The Hamiltonian for this process is given by \cite{Seifoory2017,KamandarDezfouli2017}
\begin{align}\label{eq:H_RS}
    H^\text{RS} = H^\text{RS}_\text{L} + H^\text{RS}_\text{NL},
\end{align}
where
\begin{align}
    \label{eq:LinearH_ortho}
    H^\text{RS}_\text{L} = \sum_{j = I,P,S} \hbar{\omega}_j  d^\dagger_j d_j,
\end{align}
is the linear part of the Hamiltonian, and
\begin{align}
    \label{eq:NonlinearH_ortho}
    H^\text{RS}_{\text{NL}} = \hbar \chi_\text{eff} \alpha_{P}^2(t) d^\dagger_S d^\dagger_I   + h.c.,
\end{align}
is the nonlinear part \cite{Sipe2009, Vendromin2020}. In this expression, $\alpha_{P}(t)$ is the dimensionless time-dependent coherent state amplitude of the pump \textit{in the pump mode of the RS}, which is calculated by projecting the time-evolving pump field onto the pump mode of the RS (see Appendix \ref{AppB} for details). In Eq. (\ref{eq:NonlinearH_ortho}),  
\begin{align}
    \label{eq:chi_eff}
    \chi_\text{eff} &\equiv -\frac{9\hbar}{16\epsilon_0}\sqrt{\omega_P^2\omega_S\omega_I}\nonumber\\
    &\times\sum_{i,j,k,l=1}^3 \int d^3\textbf{r}\chi^{(3)}_{ijkl}(\textbf{r})\epsilon_\text{RS}(\textbf{r})\nonumber\\
    &\times C_{P,i}(\textbf{r})C_{P,j}(\textbf{r})C^*_{S,k}(\textbf{r})C^*_{I,l}(\textbf{r})
\end{align}
is the effective nonlinear parameter carrying the dimensions of a rate, and $\chi^{(3)}_{ijkl}$ are the elements of the third-order nonlinear susceptibility tensor for the dielectric substrate, which is assumed to be homogeneous in the substrate and zero elsewhere. Using $\chi_{iiii}^{(3)}=4n_2n^2\epsilon_0 c /3$ with $n_2=4.5\times10^{-18}\ \text{m}^2\cdot \text{W}^{-1}$ for silicon at 1550 nm \cite{Dinu2003, Bristow2007}, and setting all other components to zero, we find that $\chi_\text{eff}\approx (1.84+0.07i)\times10^{5}\ \text{s}^{-1}$.

The dynamics of the generated light in the RS described by the Hamiltonian of Eq. (\ref{eq:H_RS}) is obtained by solving the Lindblad master equation for the density operator projected onto the RS QMs, $\rho^\text{RS}$ \cite{Dignam2012}. The Lindblad master equation takes the form \cite{Dignam2012} 
\begin{align}
    \label{eq:Lindblad_RS}
    &\dv{\rho^\text{RS}}{t}=-\frac{i}{\hbar}\qty[H^\text{RS}_\text{L},\rho^\text{RS}]\\&+i\sum_{j=I,S}\Gamma_j\qty(d_j\rho^\text{RS}d_j^\dagger -\frac{1}{2}d^\dagger_j d_j \rho^\text{RS}-\frac{1}{2}\rho^\text{RS}b^\dagger_j b_j),\nonumber
\end{align}
where the $\Gamma_j$ are the \textit{loaded} power decay rates of the signal $(j=S)$ and idler $(j=I)$ modes (\textit{i.e.}, the decay rates when the RS is coupled to all three CROWs), and we assume that the RS QMs are orthogonal, which is an excellent approximation for our RS. 

Vendromin and Dignam have shown that the solution to Eq. (\ref{eq:Lindblad_RS}) is the density operator of a two-mode STS \cite{Vendromin2020}, which takes the form
\begin{align}
    \label{eq:2MSTS_densityOp}
    \rho^{RS}=S_2(\xi)\rho_{th}(t)  S_2^\dagger(\xi),
\end{align}
where 
\begin{align}
\rho_{th}(t)\equiv \prod_{j=S,I} \qty(\frac{1}{n_j^\text{th}+1})\qty(\frac{n_j^\text{th}}{n_j^\text{th}+1})^{d_j^\dagger d_j}
\end{align}
is a two-mode thermal state with time-dependent thermal populations $n^{th}_j(t)$ and
\begin{align}
    \label{eq:2MSqOp}
 S_2(\xi ) = \exp(\xi^*d_S d_I - \xi d_S^\dagger d_I^\dagger)   
\end{align}
is the two-mode squeezing operator with time-dependent squeezing parameter  $\xi(t)=r(t)e^{i\theta(t)}$, where $r(t)$ is the squeezing amplitude and $\theta(t)$ is the squeezing phase. These parameters are computed by numerically solving a system of three coupled differential equations as described in Appendix \ref{AppC}. 

The $\Gamma_j$ are computed by starting the system in a single-photon Fock state in one of the RS QMs, expanded in the basis of the full-system QMs, and fitting the calculated time evolution of the photon number in the chosen RS QM to an exponential. From this procedure, we find that the loaded decay rate of the pump RS QM, which was not altered with a coupling cavity, is $\Gamma_P\approx 5.30\times10^{-5}$ $2\pi c/a$, and the intrinsic decay rate of the pump RS QM is $\Gamma_P'\approx 1.63\times10^{-5}$ $2\pi c/a$. The loaded decay rates of the signal and idler RS QMs are $\Gamma_S\approx 2.58\times10^{-5}$ and $\Gamma_I\approx 2.60\times10^{-5}$ $2\pi/c$ respectively, while their intrinsic decay rates are $\Gamma_S'\approx 7.37\times10^{-6}$ and $\Gamma_I'\approx 2.38\times10^{-5}$  $2\pi c/a$ respectively. The loaded decay rates of the signal and idler RS QMs were tuned to be approximately equivalent and close to half the loaded decay rate of the pump RS QM. Tuning was accomplished by introducing a coupling cavity (labeled in Fig. \ref{fig:PCCCS}) between the RS and each output CROW. These coupling cavities have the effect of decreasing the coupling between the RS and their respective CROWs - in this way, we decreased the couplings independently until they were at the desired values.

\begin{figure}
    \centering
    \includegraphics[width=0.48\textwidth]{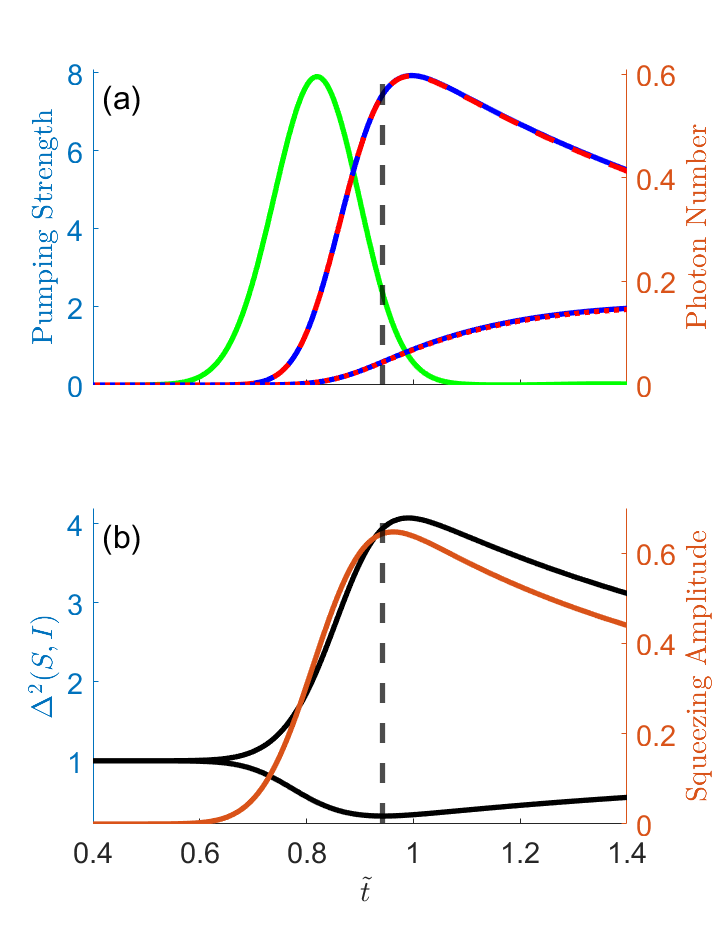}
    \caption{\justifying The time evolution of the pump strength and some key parmetrs of the squeezed thermal state in the RS for a 670 fJ Gaussian pump pulse with a $k$-space width of $\kappa=2\pi/20d$. (a) Pump strength $g(\tilde{t})$ (solid green, left axis) and photon number in the signal and idler RS QMs (right axis). The upper curves are the total photon number in the signal (blue, solid) and idler (red, dashed) RS QMs. The lower curves are the thermal photon numbers in the signal (blue, solid) and idler (red, dotted) RS QMs. Note that the results are almost identical for the signal an idler modes.   (b) Squeezing amplitude (orange, right axis) and correlation variance envelope (black, left axis) between the signal and idler RS QMs.} 
    \label{fig:gp}
\end{figure}

In all that follows, it is convenient to use the dimensionless time,
\begin{equation}
    \Tilde{t}\equiv \Gamma_+t,
\end{equation}
where 
\begin{equation}
    \Gamma_+ \equiv (\Gamma_S + \Gamma_I)/2
\end{equation}
is the average loaded loss rate for the signal and idler from the RS. To quantify the pump level in the RS, it is useful to define the pumping strength,
\begin{align}
    \label{eq:pumpingStrength}
    g(\Tilde{t}) \equiv \frac{2\abs{\alpha_P^2(\Tilde{t})\chi_\text{eff}}}{\Gamma_+},
\end{align}
which is essentially the ratio of intensity buildup of the light in the pump mode to the decay rate, $\Gamma_+$, of generated light.

In Fig. \ref{fig:gp} (a), we plot the calculated time evolution of the pump strength in the RS, $g(\Tilde{t})$ (solid green), when a Gaussian pump pulse with energy of 670 fJ, a center frequency at $\omega_P$, and $k$-space width parameter of $\kappa=2\pi/20d$ is initiated in the pump CROW (see Appendix \ref{AppB} for the expansion coefficients).
For this pump, the peak number of pump photons in the RS is  $\approx 2.3\times10^6$. In Fig. \ref{fig:gp} (a), we also show the calculated time evolution of the total photon number in the signal (red dashed) and idler (blue solid) RS QMs, and the calculated thermal photon number in the signal (red dotted) and idler (blue solid) RS QMs. We see that most of the generated photons are contributing to squeezing rather than to thermal noise. We note that the expectation value of the photon number in the generated entangled two-mode squeezed thermal state (2MSTS) is less than one, indicating the undepleted pump approximation is justified.

The correlation variance (CV) between the signal and idler RS QMs is defined as
\begin{align}
    \label{eq:RSCV_general}
    \Delta_{S,I}^2&=\expval{\qty[\Delta\qty(X_S-X_I)]^2}\nonumber\\ & + \expval{\qty[\Delta\qty(Y_S+Y_I)]^2},
\end{align}
where
\begin{align}
    X_j &= \frac{1}{2}\qty(d_j e^{i\theta_j} + d_j^\dagger e^{-i\theta_j}),\label{eq:X_j}\\
    Y_j &= \frac{1}{2i}\qty(d_j e^{i\theta_j}- d_j^\dagger e^{-i\theta_j}),\label{eq:Y_j}
\end{align}
are the quadrature operators for the $j^\text{th}$ RS QM.  In these expressions,  $\theta_S(t)$ and $\theta_I(t)$ are time-dependent phases that can be chosen as desired. The correlation variance is an indicator of inseparability: if $\Delta^2_{S,I} < 1$, then the state is entangled \cite{Duan2000}.
After substituting Eqs. (\ref{eq:X_j}, \ref{eq:Y_j}) into Eq. (\ref{eq:RSCV_general}) we obtain

\begin{align}
    \Delta^2_{S,I}&=\qty(n_S^\text{th}+n_I^\text{th}+1)\\
    &\times\qty{\cosh(2r)+\cos(\theta_S+\theta_I+\theta)\sinh(2r)}\nonumber.
\end{align}

To isolate the minima (maxima) of the CV we set the argument of the second cosine to $\pi$ $(0)$ by selecting appropriate $\theta_S(t)$ and $\theta_I(t)$ such that $\theta_S + \theta_I + \theta = \pi\ (0)$. We thus obtain expressions for the maximum and minimum CV in the between the signal and idler modes of the RS:
\begin{align}
     \Delta_{S,I;\text{min}}^2 &= \qty(n_S^\text{th}+n_I^\text{th}+1)e^{-2r},\\
     \Delta_{S,I;\text{max}}^2 &= \qty(n_S^\text{th}+n_I^\text{th}+1)e^{2r}.
\end{align}

In Fig. \ref{fig:gp} (b), we plot the time evolution of the squeezing amplitude, $r(\Tilde{t})$ (orange), and the minium and maximum of the correlation variance in the RS (black). As can be seen, the CV reaches a minimum value of $\approx0.2975$ at time $\Tilde{t}_0\approx 0.9428$. We retrieve the calculated quantities $r(\Tilde{t}_0)$, $n_S^\text{th}(\Tilde{t}_0)$, and $n_I^\text{th}(\Tilde{t}_0)$, given in Table \ref{tab:genStateParams}, to define the density operator of the generated two-mode squeezed thermal state at the time $\Tilde{t}_0$ according to Eq. (\ref{eq:2MSTS_densityOp}). Note that, because we have designed the structure such that the signal and idler decay times are nearly equal and are much longer than the pump pulse duration in the RS, the number of thermal photons is very low and is much smaller than the total number of photons. The properties of the 2MSTS in the RS at $\Tilde{t}_0$ are summarized in Table \ref{tab:genStateParams}. 

In the next section, we use this density operator as the starting point to study the properties of the generated light in the signal and idler CROWs for times $\Tilde{t}\geq\Tilde{t}_0$. 

\begin{table}[h!]
    \centering
    \begin{tabular}{c|c|c}
       \ \ $r(\Tilde{t}_0)$\ \ & \ \ $n_S^\text{th}(\Tilde{t}_0)$\ \ & 
 \ \ $n_I^\text{th}(\Tilde{t}_0)$\ \  \\ \hline
        0.6497 & 0.0458 & 0.0453\\ \hline\hline
        $\Delta_{S,I;\text{min}}^2(\Tilde{t_0})$ & $n_S(\Tilde{t_0})$ & $n_I(\Tilde{t_0})$\\\hline
        0.2975 & 0.5749 & 0.5744
    \end{tabular}
    \caption{Summary of parameters for the state generated in the RS at time $\Tilde{t}=\Tilde{t}_0$.}
    \label{tab:genStateParams}
\end{table}

\subsection{Quantum Evolution}
In this subsection we present our approach to calculating the propagation of the light generated in the RS throughout the full system and into the two output CROWs. We start by assuming that at time $\tilde{t}=\tilde{t}_0$, the system is in the density operator $\rho(\tilde{t}_0)$ calculated in the previous subsection. We then assume that the effect of the pump is small for $\Tilde{t}\geq\Tilde{t}_0$ and make a key approximation that we can set it to zero for these times. 

The neglect of the pump light greatly simplifies the calculation of the time evolution throughout the system because simulating the nonlinear interaction in the basis of full-system QMs is essentially computationally intractable. First of all, one does not know \textit{a priori} which QMs will contribute significantly to the dynamics, requiring that one must include all QMs in the simulation. In addition, there is no known solution to the problem of multimode DFWM in a lossy system if there are more than two modes, meaning the problem has to be solved numerically. As discussed in the introduction, modeling our system with a Fock basis approach - even if we only include in our basis the 33 QM states that have 1\% or more of the photon population at the starting time $\Tilde{t}_0$ - would require more than $5\times10^{15}$ basis states. However, if we assume that the pump is negligible for $\Tilde{t}\geq\Tilde{t}_0$, then the generated 2MSTS evolves under the free Hamiltonian only, and we can calculate this evolution analytically in the basis of full-system QMs. Moreover, it can be shown that the assumption that the pump vanishes at time $\Tilde{t}_0$ is a conservative approximation: the resulting correlation variance is guaranteed to be larger (worse) than that obtained with the pump present \cite{Dignam2025}.

We now present the exact solution to the adjoint master equation for the free evolution of the full-system QM ladder operators (See Appendix \ref{AppD} for details). 
Due to the non-orthogonality of the basis of full-system QMs the non-Hermitian free Hamiltonian describing light in the full system takes the form \cite{Dignam2012}
\begin{align}
    \label{eq:H_free}
    H_\text{free} = \sum_{\mu=1}^N\hbar\comega_\mu b^\dagger_\mu c_\mu,
\end{align}
where $N=184$ is the total number of full-system QMs (equal to the number of defect cavities), and the full-system QM ladder operators obey the following relations:
\begin{align}
    \label{eq:bcrelations}
    &c_\mu=\sum_\nu P_{\mu\nu}b_\nu,\quad b_\mu=\sum_\nu O_{\mu\nu}c_\nu,\\
    &c_\mu^\dagger = \sum_\nu P_{\mu\nu}^*b_\nu^\dagger,\quad b^\dagger_\mu  = \sum_\nu O_{\mu\nu}^*c_\nu^\dagger,
\end{align}
where $\mat{O}$ the overlap matrix for the full system QMs, with elements
\begin{equation}
     O_{\mu\nu}\equiv \int_\infty d^3 \textbf{r}\epsilon(\textbf{r}) \textbf{N}_\mu^*(\textbf{r})\cdot\textbf{N}_\nu(\textbf{r})
\end{equation}
and $\mat{P}\equiv\mat{O}^{-1}$. The full-system QM ladder operators obey the commutation relations:
\begin{align}
    &[c_\mu,\ b_\mu^\dagger]=\delta_{\mu\nu},\\
    &[b_\mu,\ c_\nu^\dagger]=-\delta_{\mu\nu},\\
    &[b_\mu,\ b_\nu^\dagger]=O_{\mu\nu},\\
    &[b_\mu^\dagger,\ b_\nu^\dagger]=[c_\mu,\ c_\nu]=0.
\end{align}
The state with a single photon in the $\mu^\text{th}$ full-system QM is created by the rightwards action of $b^\dagger$ on the vacuum, which is an eigenstate of the linear Hamiltonian of Eq. (\ref{eq:H_free}). We can construct the full-system QM ladder operators in terms of the single-cavity QM ladder operators following Eq. (\ref{eq:NmodeExp}):
\begin{align}
    \label{eq:b_over_a}
    b_\mu^\dagger = \sum_q v_{q\mu}a_q^\dagger,\quad b_\mu=\sum_q v_{q\mu}^*a_q.
\end{align}
Inverting Eq. (\ref{eq:b_over_a}) yields the inverse relations
\begin{align}
    \label{eq:single_cavity_inverse_tb}
    a_q^\dagger = \sum_\mu w_{\mu q}b_\mu^\dagger,\quad a_q=\sum_\mu w_{\mu q}^*b_\mu.
\end{align}

Solving the adjoint master equation for the time-evolution of the full-system QM ladder operators (see Appendix D) gives the simple result: 
\begin{align}
    \label{eq:TE_c}
c_\nu(\Tilde{t}) = c_\nu e^{-i\Tilde{\omega}_\nu (\Tilde{t}-\tilde{t}_0)},
\end{align}
where $c_\nu\equiv c_\nu(\Tilde{t}_0)$. For $c^\dagger(\Tilde{t})$ we simply take the Hermitian conjugate of Eq. (\ref{eq:TE_c}). More generally, it is easy to show that for any normally-ordered set of $c-$operators, the evolution of each operator is simply given by the result shown in Eq. (\ref{eq:TE_c}). 
We can now use this to calculate the time evolution of the system in any of the single-defect cavities in the CROWs. The first step is to determine the time evolution of the single-defect QM ladder operators.  As shown in Appendix \ref{AppD}, this is given by:
\begin{align}    
    \vec{a}(\Tilde{t})&=\mat{W}^\dagger\mat{O}\mat{F}(\Tilde{t})\mat{P}\mat{V}^\dagger \vec{a}(\Tilde{t}_o)\\
    &\equiv\mat{\Phi}(\Tilde{t})\vec{a}(\Tilde{t}_o),\label{eq:TE_a}
\end{align}
where $\mat{W}\equiv\mat{V}^{-1}$, $\vec{a}\equiv\qty(a_1,a_2,...,a_p)^T$, and 
\begin{align}&\mat{F}(\Tilde{t})\equiv  \\
&\text{diag}\qty(e^{-i\comega_1(\Tilde{t}-\tilde{t}_0)},e^{-     i\comega_2(\Tilde{t}-\tilde{t}_0)}, ... ,e^{-i\comega_\nu (\Tilde{t}-\tilde{t}_0)}). \nonumber
\end{align}
Knowing the time evolution of the single-defect ladder operators, we can determine the expectation value of any normally-ordered combination of ladder operators, and hence we have access to all the information about the state of the light in the PC CCS. In the remainder of this paper, we will focus on the photon number in individual cavities and the CV between single-defect cavities in the signal and idler CROWs. 

Using Eq. \ref{eq:TE_a}, we obtain the following expression for the time-dependent photon number in the $p^\text{th}$ cavity:
\begin{align}
    \label{eq:TE_PN}
    \expval{a_p^\dagger a_p}(\Tilde{t}) = \sum_{q\in RS} \phi_{qp}^*(\Tilde{t})\phi_{qp}(\Tilde{t})\expval{a_q^\dagger a_q}.
\end{align}
As is shown in Appendix \ref{AppD}, the ensemble averages at $\tilde{t}=\tilde{t}_0$ are given by:
\begin{align}
    &\expval{a_p a_q} =\label{eq:aa}\\
    &-\qty({\sigma}^*_{pS} {\sigma}^*_{qI}+ {\sigma}^*_{pI} {\sigma}^*_{qS})\nonumber\\
    &\times \qty(n^{th}_S+n^{th}_I+1)e^{i\theta}\cosh r, \sinh r,\nonumber
\end{align}
and
\begin{align}
    &\expval{a_p^\dagger a_q} =\label{eq:adaga}\\
    &\sigma_{pS} {\sigma}^*_{qS} \qty(n^{th}_S\cosh^2r + (1+n^{th}_I)\sinh^2r) \nonumber\\
    &+ \sigma_{pI} {\sigma}^*_{qI} \qty(n^{th}_I\cosh^2r + (1+n^{th}_S)\sinh^2 r),\nonumber
\end{align}
where $\sigma_{pj}$ are the matrix elements of the matrix $\mat{\Sigma}=\mat{U}^{-1}$ (see Eq. (\ref{eq:TBEVP_RS}), and $r$, $n^\text{th}_S$, and $n^\text{th}_I$ are evaluated at $\tilde{t}=\tilde{t}_0$. Similarly, the CV between cavities $p_S$ in the signal CROW and $p_I$ in the idler CROW is given by:

\begin{align}
    &\Delta^2(p_S,\Tilde{t}_S;p_I,\Tilde{t}_I) =\label{eq:outputCV_simplified}\\
    &\frac{1}{2}(B_{p_{S} p_{S}} + B_{p_{I} p_{I}})\nonumber\\
    &+\sum_{q,q'=1}^N [\phi_{p_{S}q}^*(\Tilde{t}_S)\phi_{p_{S}q'}(\Tilde{t}_S)\nonumber\\
    &+ \phi_{p_{I}q}^*(\Tilde{t}_I)\phi_{p_{I}q'}(\Tilde{t}_I)]\expval{a^\dagger_q a_{q'}}\nonumber\\
    &-2\Re{\sum_{q,q'=1}^N\phi_{p_{S}q}(\Tilde{t}_s)\phi_{p_{I}q'}(\Tilde{t}_I)\expval{a_q a_{q'}}}.\nonumber
\end{align}

where $\phi_{qn}(\Tilde{t})$ are the elements of $\mat\Phi(\Tilde{t})$, $B_{pq}$ is the coupling matrix element between the $p^\text{th}$ and $q^\text{th}$ cavities (see Appendix \ref{AppA}). The full derivation of Eqs. (\ref{eq:TE_PN}) and (\ref{eq:outputCV_simplified}) are given in Appendix D. 

\section{Results}\label{sec:IV}
In this section we analyze the generated light from the PC CCS shown in Fig. \ref{fig:PCCCS} using our theoretical approach. In particular, we present the time evolution of the average photon number and CV in cavities in the three CROWs. We examine the sensitivity of the CV to changes in pump energy and spectral width. We begin by examining the time-dependent average photon number in cavities in the three CROWs to ascertain the spatio-temporal dynamics of the generated light. Next we compute the time-dependent CV between light in the signal and idler CROWs by solving Eq. (\ref{eq:outputCV_simplified}). We will explore a variety of pumping configurations and time delays to demonstrate the utility of our approach and the performance of our coupled cavity system. 
\subsection{Photon Number}
We begin by using Eq. (\ref{eq:TE_PN}) to calculate the time dependence of the photon number for the first few cavities in each of the three CROWs for the pump pulse that was used to produce the results presented in Fig. \ref{fig:gp}. The photons we study when computing the time-dependent photon number here are the generated photons, which exist in a multimode entangled squeezed thermal state in the QMs of the full system. 

\begin{figure}[htp]
    \centering
    \includegraphics[width=0.48\textwidth]{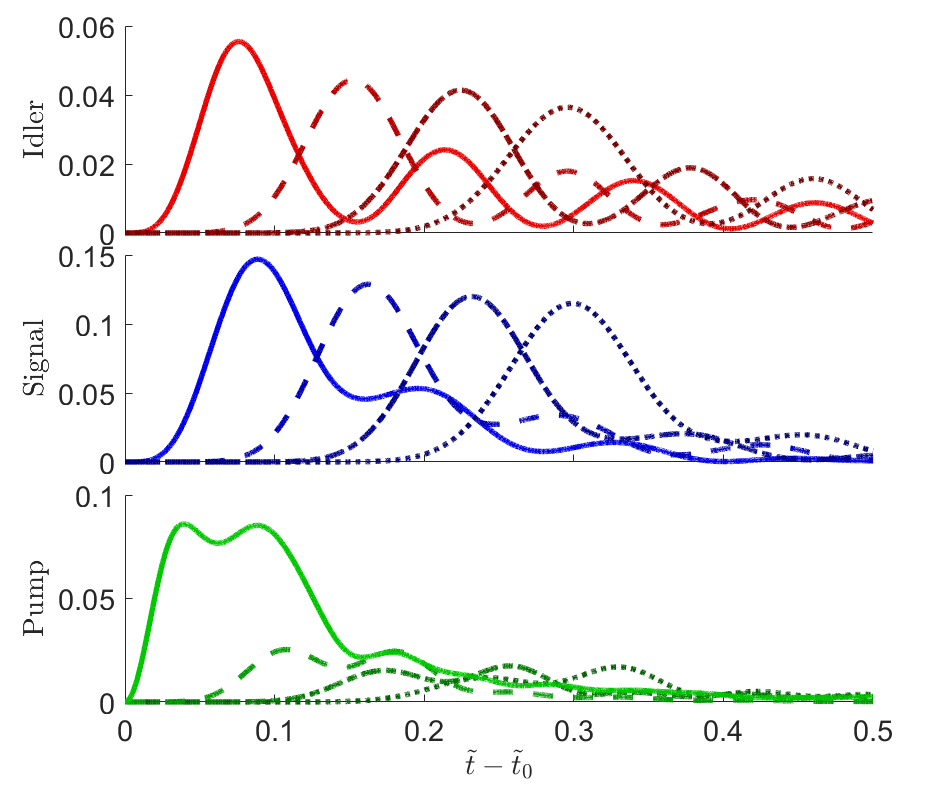}
    \caption{\justifying Average photon number in the $1^\text{st}$ (solid), $4^\text{th}$ (dash), $7^\text{th}$ (dash-dot), and $10^\text{th}$ (dot) cavities of the idler, signal, and pump CROWs as a function of time. The labeling of the cavities is given in Fig. \ref{fig:PCCCS}. Note that the photon numbers do not include the input pump pulse photons, but rather the  photons generated in the RS.}
    \label{fig:outputCROW_PN}
\end{figure}

In Fig. \ref{fig:outputCROW_PN}, we plot the photon number in select cavities in the three CROWs as a function of time. In the signal and idler CROWS, the photon number in a given cavity exhibits a series of peaks that decrease in amplitude with time. The time that the first peak arrives in each cavity increases, and its amplitude decreases as the distance of the cavity from the RS increases. This is simply due to the effects of the group velocity and scattering loss, respectively. One might expect that the period between successive peaks in the CROWS to be given by the beat period between the signal and idler modes in the RS.  This is approximately the case in the pump CROW. However, complications arising from reflections off the CROWS and the coupling cavities results in the oscillation period in the signal and idler CROWS being closer to twice the beat period.

\begin{figure}[htp]
    \centering
    \includegraphics[width=0.48\textwidth]{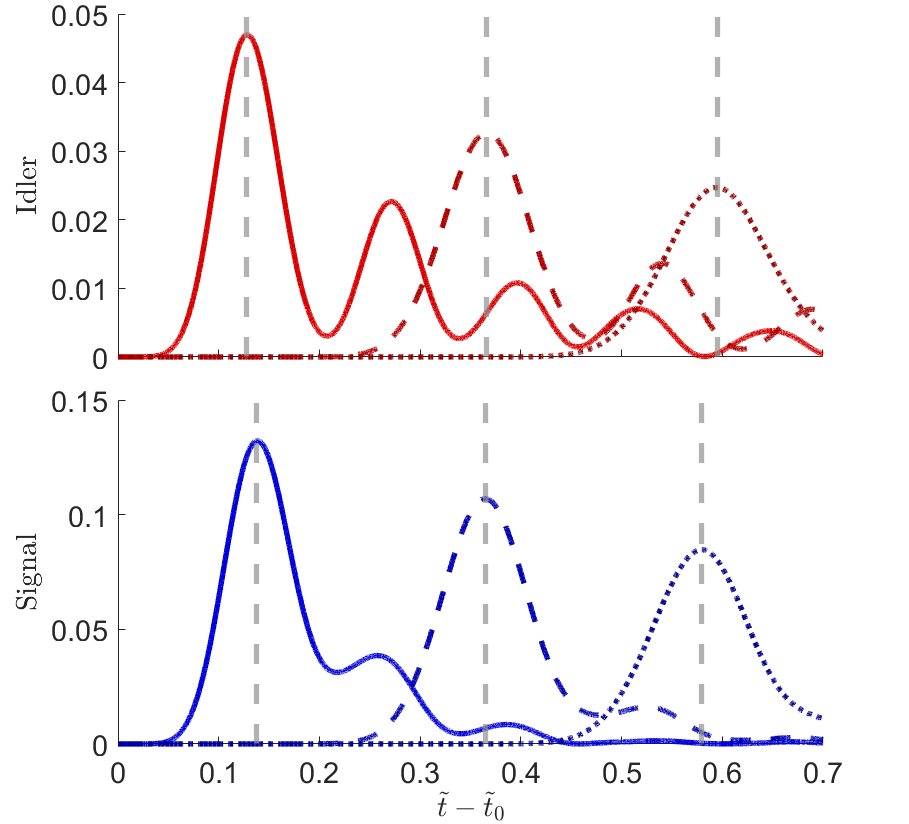}
    \caption{\justifying Photon numbers in the $3^\text{rd}$ (solid), $13^\text{th}$ (dashed), and $23^\text{rd}$ (dotted) signal and idler CROW cavities as a function of time. Note that the signal light overtakes the idler when they reach the $13^\text{th}$ cavities in the CROWs. Vertical dashed lines indicate the pulse pulse peaks. }
    \label{fig:PN_racing}
\end{figure}

Another feature seen in Fig. \ref{fig:outputCROW_PN}, which is pertinent to the following analysis of the CROW CV, is the difference in the propagation speeds in the signal and idler CROWs. Each of the CROWs is designed to carry light at its band center. Due to the difference in inter-cavity coupling in each of the CROWs, the group velocity at band center is different in each CROW. We infer this from Fig. \ref{fig:crowbands}, where the bandwidths of the three CROWs are different, and so the group velocities at band center are different. Indeed, we see that the average photon number peaks earlier in the first idler cavity than the first signal cavity, though by cavity 13 the signal light begins arriving first. This phenomenon is clearly seen in Fig. \ref{fig:PN_racing}, where we plot the photon number in three different cavities in the signal and idler CROWS as a function of time. 

From Fig. \ref{fig:outputCROW_PN}, we also see that for cavities sufficiently far from the RS, most of the generated light is in the signal and idler waveguides, as desired.
At early times however, a substantial fraction of the generated light is in the pump CROW cavities close to the RS. Most of this light cannot propagate down the pump CROW because its frequency is outside of the pass-band of the pump CROW (see Fig. \ref{fig:crowbands}). Consequently, it evanescently decays and its amplitude is very small beyond the third cavity in the pump CROW. The relatively large number of generated photons in the cavities of the pump CROW that are close to the RS arises due to the strong coupling of the pump CROW to the RS. We have deliberately chosen this coupling to be strong to reduce the residence time of the pump light in the RS; in studying this system and others similar to it, we have found that a shorter residence time limits the generation of thermal noise from scattering loss during the nonlinear process in the RS. 

Finally, we observe from Fig. \ref{fig:outputCROW_PN} that that there are more photons in the signal CROW than in the idler CROW. There are two different contributors to this imbalance. The first is the result of the different coupling strengths between the RS and the signal and idler CROWs. Since the intrinsic Qs of the signal and idler RS QMs were different, the coupling between these QMs and their respective CROWs had to be tuned separately to achieve the same loaded $Q$ for the signal and idler RS modes. This was accomplished via the coupling cavities indicated in Fig. \ref{fig:PCCCS}. Manipulating the loaded $Q$s of the RS QMs in this manner resulted in a difference in the out-coupling rate between the signal and idler modes. Ideally the loaded $Q$s and the out-coupling efficiency of the signal and idler RS QMs would be approximately equal, but it was not possible to achieve this for this structure, while also keeping the equal frequency spacing between the modes. The second contributor to the imbalance in the signal and idler photon number is the effect of the strongly-coupled pump CROW, which results in a more complicated time evolution of the generated light after the pump has left. Indeed, when we perform the same analysis but consider a system with the pump CROW removed, the time-dependent photon numbers in the signal and idler CROWs are nearly equal. 

\subsection{Correlation Variance}
\begin{figure}[htp]
    \centering
    \includegraphics[width=0.48\textwidth]{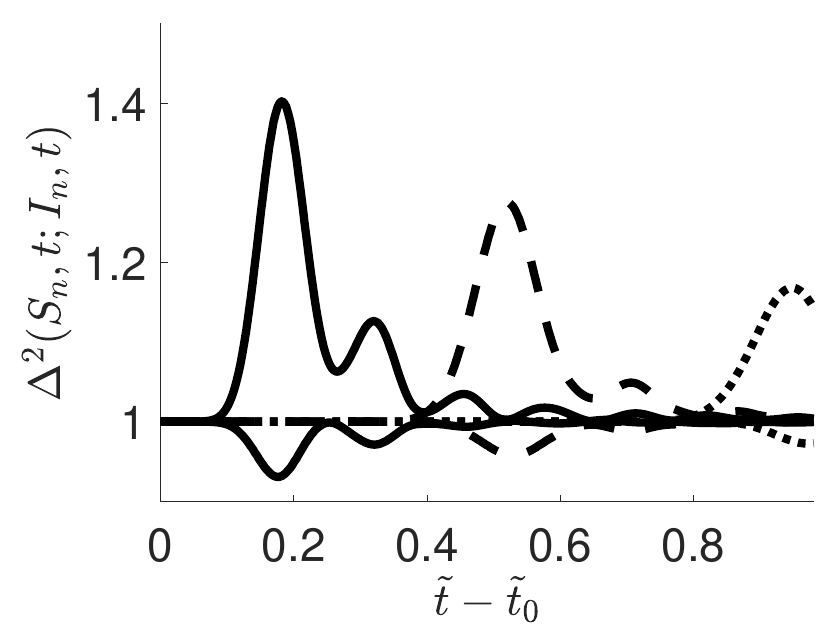}
    \caption{\justifying Envelope of the CV between pairs of signal and idler CROW cavities as a function of time. The fast oscillations that arise in the CV have been omitted for clarity and only the envelope has been plotted. Curves shown are for the $5^\text{th}$ (solid), $20^\text{th}$ (dashed), and $40^\text{th}$ (dotted) pairs away from the coupling cavities. }
    \label{fig:CV_cavity_pairs}
\end{figure}

We now turn our attention to evaluating the correlation variance between signal and idler CROW cavities as expressed in Eq. (\ref{eq:outputCV_simplified}). We once again use the pulse shown in Fig. \ref{fig:gp}. For simplicity, we first set $\Tilde{t}_S=\Tilde{t}_I$ in Eq. (\ref{eq:outputCV_simplified}). The CV expression in Eq. (\ref{eq:outputCV_simplified}) includes rapidly oscillating terms modulated by a slowly varying envelope. Experimentally, these fast oscillations will be removed during a heterodyne measurement by locking the measurement phase with the local oscillators. Thus for simplicity in Fig. \ref{fig:CV_cavity_pairs} we plot only the slowly varying upper and lower bounding envelopes of the CV function. The lower bound of the envelope is a measure of inseparability of the state: the state is inseparable if this lower bound of the CV is less than unity. We see from Fig. \ref{fig:CV_cavity_pairs} that in the fifth cavity pair, the CV falls to about 0.93. Moreover, entanglement between pairs of cavities persists for cavities that are more than 40$d$ = 57.6 $\mu$m away from the coupling cavities. 

\begin{figure}[htp]
    \centering
    \includegraphics[width=0.48\textwidth]{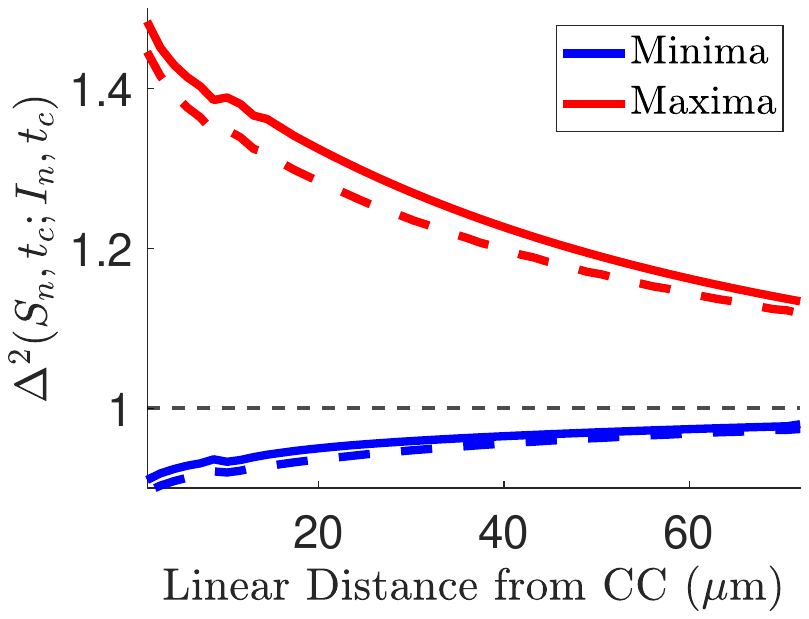}
    \caption{\justifying Minimum and maximum of the CV envelope for pairs of output CROW cavities as a function of distance the coupling cavity (CC). Dashed lines show the results obtained assuming that the system QMs are orthogonal. }
    \label{fig:CV_extrema}
\end{figure}

In Fig. \ref{fig:CV_extrema}, we plot the global minima and maxima of the CV envelopes from the first 50 output CROW cavity pairs, with $\Tilde{t}_S=\Tilde{t}_I$. The minima of the CV remain below unity in all cases, demonstrating entanglement far from the RS. 

We now examine the role that the nonorthogonality of the full system QMs plays in the state dynamics. As shown in Fig. \ref{fig:FullSystemOverlaps}, there are many modes that have very significant overlap ($\vert O_{\mu,\nu}\vert > 0.1$). To quantify the effect of this nonorthogonality, 
we have repeated the calculation of the CV ignoring the non-orthogonality, \textit{i.e.}, setting $\mat{O}=\mat{I}$ in Eq. (\ref{eq:TE_a}), and present the results for the CV envelope as the dashed lines in Fig. \ref{fig:CV_extrema}. Note that for cavities close to the RS, the difference between the CV minimum and unity can change by up to 16\%, indicating that it is important to include the nonorthogonality of the modes in the calculations for this system.

We now consider the effects of varying the pump pulse energy and width in $k$-space on the CV and the optimal measurement times. Although the na\"ive choice of $\Tilde{t}_S=\Tilde{t}_I$ in Eq. (\ref{eq:outputCV_simplified}) is satisfactory for low pump energies, the increased thermal noise that is generated as the pump energy increases eventually forces one to alter the measurement times to account for the group velocity difference. To demonstrate this, we calculated the CV between the $n=3$ cavities in the signal and idler CROWS under a variety of pumping configurations and measurement times.  In what follows, we fix $\Tilde{t}_S$ and define the measurement delay time as $\Tilde{\tau}=\Tilde{t}_S-\Tilde{t}_I$. 

\begin{figure}[htp]
    \centering
    \includegraphics[width=0.48\textwidth]{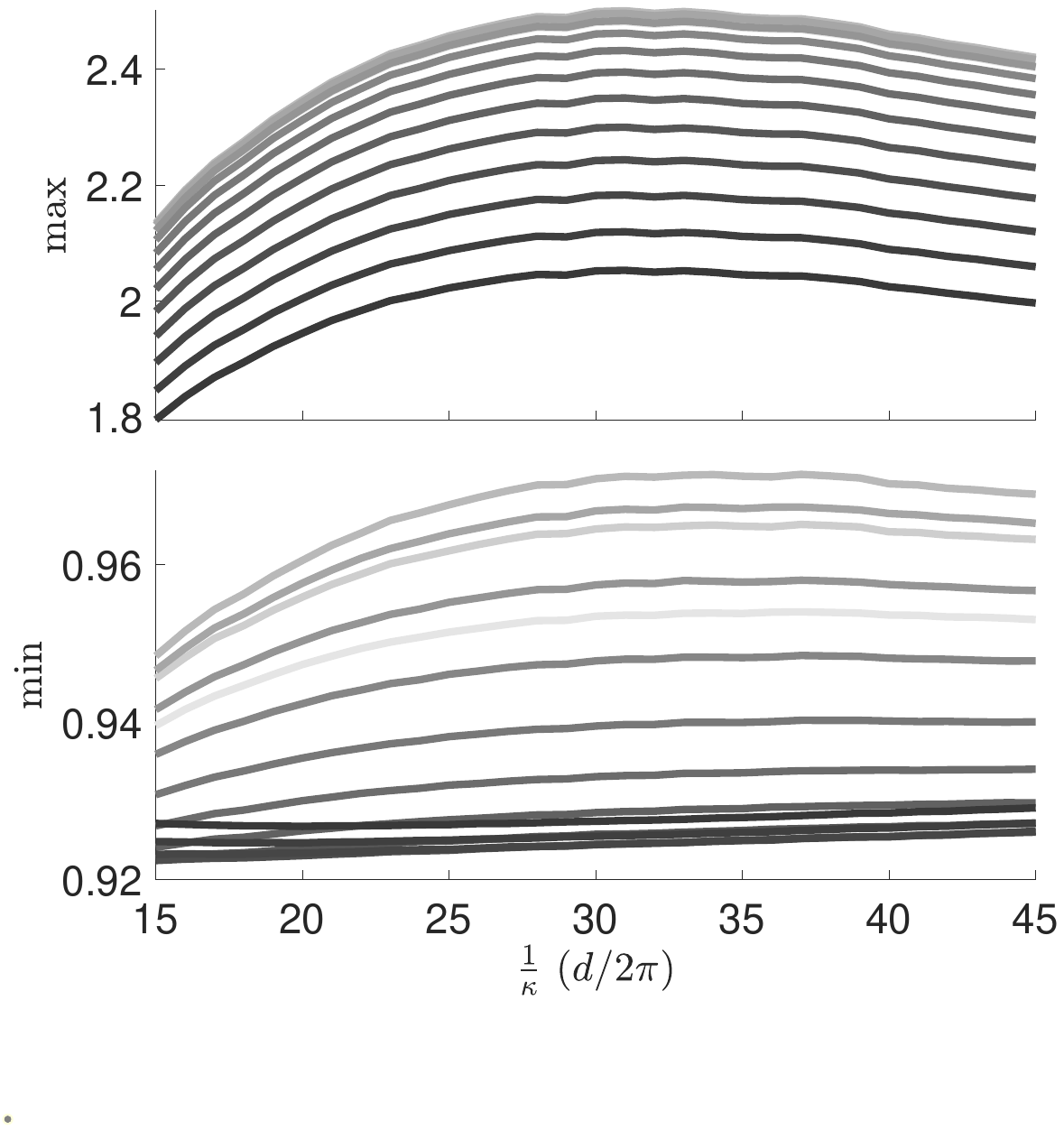}
    \caption{\justifying Minimum and maximum of the CV for the $n=3$ signal and idler cavities as a function of $1/\kappa$, with a fixed pump energy of 670 fJ. Each curve corresponds to a particular choice of $\Tilde{\tau}$ from $\Tilde{\tau}=0$ (lightest curve) to $\Tilde{\tau}=0.065\Gamma_+$ (darkest curve) in increments of $0.005\Gamma_+$.}
    \label{fig:kappa_sweep}
\end{figure}

In Fig. \ref{fig:kappa_sweep}, we plot the minimum and maximum CV as a function of $\kappa^{-1}$ for a fixed pump energy of 670 fJ for fourteen different $\tilde{\tau}$, from $\Tilde{\tau}=0$ to $\Tilde{\tau} =0.065\Gamma_+$ in increments of $0.005\Gamma_+$. Note that the minimum CV initially increases and then decreases as $\tilde{\tau}$ is increased, while the maximum CV strictly decreases.  The effect of varying $\kappa$ between $2\pi/15d$ and $2\pi/45d$ is less pronounced than that of varying the measurement time delay $\Tilde{\tau}$. Specifically, with a fixed pump energy of 670 fJ (as in Fig. \ref{fig:kappa_sweep}), we find that choosing $0.045\Gamma_+\leq\tilde{\tau}\leq0.065\Gamma_+$ (the darkest six curves) greatly reduces the dependence of the CV minimum on $1/\kappa$, while also reducing the CV maximum. From an alternative perspective, if the optimum $\Tilde{\tau}$ is chosen to minimize the CV minimum, then the frequency width (and thus the temporal width) of the pump pulse may be adjusted to minimize the CV maximum, which has little effect on the CV minimum. 

\begin{figure}[htp]
    \centering
    \includegraphics[width=0.48\textwidth]{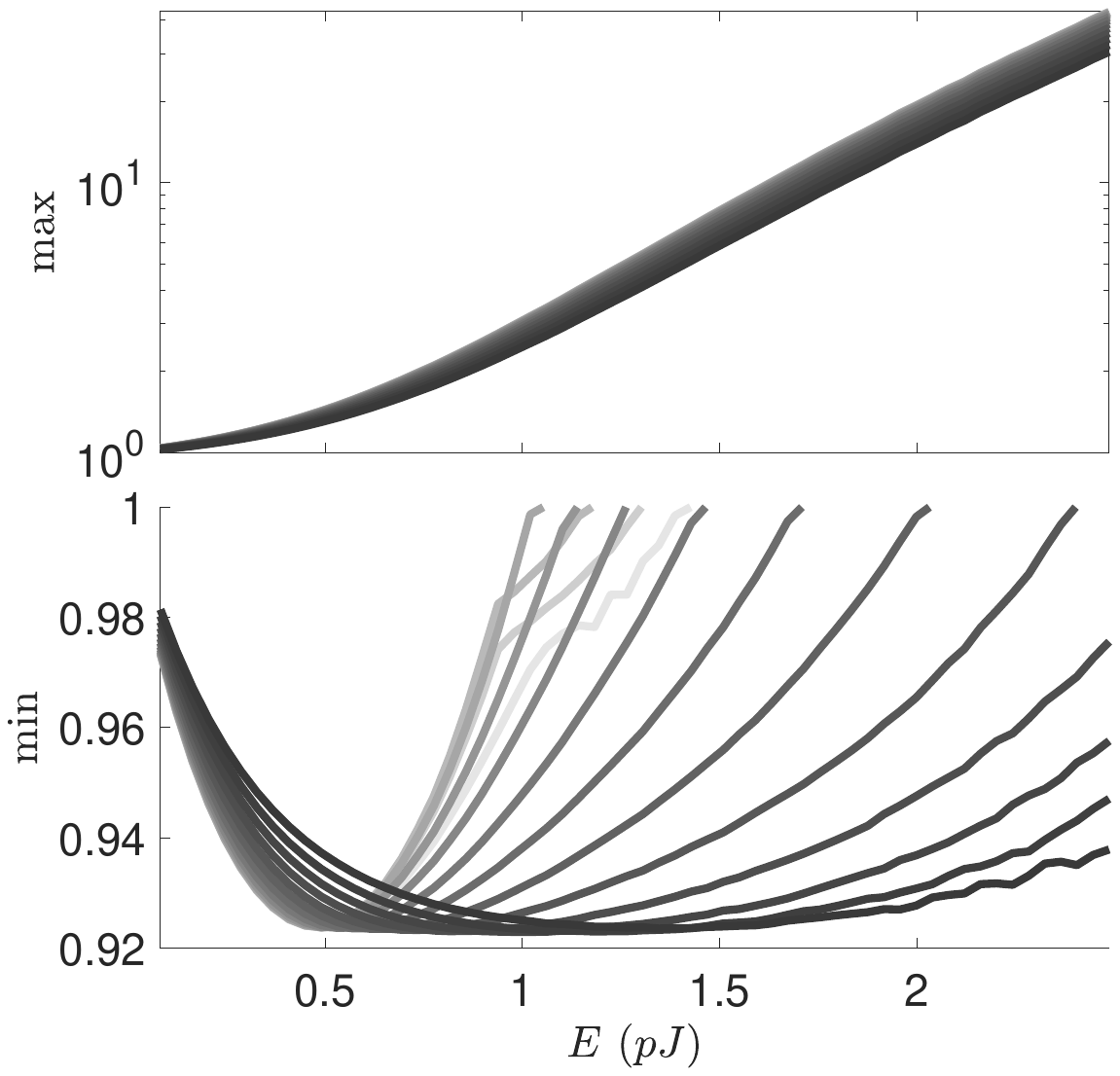}
    \caption{\justifying CV for the $n=3$ signal and idler cavities as a function of pump energy with $\kappa = 2\pi/20d$ fixed. Each curve corresponds to a particular choice of $\Tilde{\tau}$ from $\Tilde{\tau}=0$ (lightest curve) to $\Tilde{\tau}=0.065\Gamma_+$ (darkest curve) in increments of $0.005\Gamma_+$.}
    \label{fig:E_sweep}
\end{figure}

In Fig. \ref{fig:E_sweep}, we plot the minimum and maximum CV as a function of pump energy for a fixed $\kappa=2\pi/20d$. The pump energy has a more pronounced effect on the CV extrema than $\kappa$. Fig. \ref{fig:E_sweep} (b) shows that once the pump energy exceeds $\approx1$ pJ, the entanglement vanishes unless the time delay is sufficiently increased \footnote{The jogs in the CV minima in the lighter curves are caused by secondary peaks becoming more inseparable than the initial pulse for high enough pump energy.}. From Fig. \ref{fig:E_sweep} (a), we see that for pulse energies above $\approx 1$ pJ, the CV maximum exhibits an exponential dependence on the pump energy: although entanglement may be recovered with an appropriate choice of $\Tilde{\tau}$, the large increase in the CV maximum makes these higher pulse energies undesirable. We find that a good configuration that gives a low minimum CV, while keeping the pulse energy and the maximum CV modest, is as follows: a pump pulse energy of $\approx 670$ fJ, $\kappa=2\pi/20d$, and $\Tilde{\tau}=0.45\Gamma_+$. This gives a minimum and maximum CV for the $n=3$ cavity pair of 0.923 and 1.74 respectively.

\section{Conclusion}\label{sec:con}
In this work we modeled the generation and propagation of a 2MSTS in a PC CCS with 184 cavities. We studied the impact of loss and QM non-orthogonality on the evolution of the generated light, and we investigated the effects of the properties of the PC CCS on the generation efficacy of the system. The key to our theoretical approach is that we separate the problem into 
four separate, tractable steps. This makes it possible to simulate quantum state generation and propagation in large, multimode, coupled cavity systems that cannot be simulated using conventional Fock-basis approaches. 

Our analysis showed that the RS, when excited with a Gaussian pump pulse carrying an initial energy between 0.5 and 1 pJ and a $k$-space width between $2\pi/15 d$ and $2\pi/45d$, generates a two-mode squeezed thermal state, which exits the RS as two spatially separated pulses of light that remain entangled as they propagate. We find that although the nonlinear generation process creates strong squeezing and low thermal photon noise in the RS, the squeezing degrades substantially after transmission into the output CROWs. The difference between the intrinsic quality factors of the RS QMs and the over-coupling of the pump CROW contribute to this degradation. Thus, it is clear that further refinement of the design of the RS and coupling to the CROWs in this system is required to balance the number of photons going into each of the CROWs, while keeping the loaded cavity Qs in the desired range. 



Although we focused on the time-dependent photon number and CV in this work, our method is general and can be applied to any observable that can be expanded in terms of the full-system QMs. The general procedure applied here may also be extended to any nanophotonic system composed of coupled elements whose modes are easily obtained. 

Possible extensions of the theoretical approach presented here include the handling of other device types, such as line-defect waveguides, and a relaxation of the undepleted pump approximation. Other applications of the theoretical approach presented here include designing structures for the on-chip generation of larger numbers of entangled states, including cluster states and non-Gaussian states, designing implementations of quantum gates on-chip, and studying the dynamics of quantum information protocols in the presence of loss.



\begin{acknowledgments}
This work was supported by The Canadian Foundation for Innovation and the Natural Sciences and Engineering Research Council of Canada. 
\end{acknowledgments}


\newpage
\bibliography{Refs}

\appendix

\section{Quasi-Mode Bases}\label{AppA}
In this appendix we describe in detail the calculation of the various bases used in our theoretical approach. As our first step, we calculate the QMs of the individual defects and  present the tight-binding formalism that we use to obtain the leaky QMs of the entire PC CCS discussed earlier and shown in Fig. \ref{fig:PCCCS}. We first discuss some general properties of PC slabs and defect modes, and then show how we determine the QMs of a general system of coupled defects. We then compute the QMs of the full system and each of the subsystems. 
\subsubsection{General Procedure}\label{sec:IIIA1}
Defect-free PC slabs are periodic dielectric structures which, if properly designed, can support a photonic band gap - a set of frequencies for which light cannot propagate in the PC - similar to an electronic band gap in which no electronic states exist \cite{PCBook}. Both of these band gaps arise from the periodicity of the underlying structure. The photonic band structure (and thus the size and location of the photonic band gap) of a PC slab is easily calculated using finite-difference time-domain (FDTD) simulations or finite-element analysis (FEA) by exploiting the high symmetry in a defect-free PC slab. 

One may form \textit{single-defect cavities} in a PC slab by removing or shrinking an air hole at a particular lattice site, and thus locally disrupting the periodicity of the structure. Defects of this type are displayed in Fig. \ref{fig:PCCCS}. These single-defect cavities in a square-lattice PC slab support a quadrupole-like resonant QM localized to the lattice site of the altered air hole \cite{Fussell2007}. The resonant frequency of the single-defect cavity modes formed in this manner lies within the transverse electric photonic band gap of the underlying PC slab. 

In principle, one could obtain the QMs of a system of coupled defects via FDTD or FEA, though achieving accurate results from these simulations becomes prohibitively resource-intensive as the number of defects increases, especially if the system has low symmetry and the modes are close together in frequency relative to their intrinsic linewidth. Instead, we obtain the system QMs through a tight-binding expansion of the system QMs in terms of the (lossy) QMs of the individual defects that comprise the PC CCS.

The positive frequency part of the electric field of light in the $q^\text{th}$ individual defect QM is given by \cite{KamandarDezfouli2014}
\begin{align}
    \label{eq:cavityQM_E+}
    \textbf{E}^+_q(\textbf{r},t) = \sqrt{\frac{\hbar\Omega_q}{2\epsilon_0}}\textbf{M}_q(\textbf{r})e^{-i\Tilde{\Omega}_q t},
\end{align}
where $\textbf{M}_q(\textbf{r})$ is the spatial profile of the leaky resonant mode supported by the $q^\text{th}$ cavity. These QMs are subject to normalization condition
\begin{equation}
    \int d^3\textbf{r}\epsilon_p(\textbf{r})\textbf{M}_p^*(\textbf{r})\cdot\textbf{M}_p(\textbf{r})=1,
\end{equation}
and satisfy the Helmholtz equation
\begin{align}\label{eq:helmholtz_iso}
    \curl\curl\textbf{M}_q(\textbf{r}) - \frac{\Tilde{\Omega}_q^2}{c^2}\epsilon_q(\textbf{r})\textbf{M}_q(\textbf{r}) = 0,
\end{align}
where $\epsilon_q(\textbf{r})$ is the dielectric function of the PC containing only the $q^\text{th}$ defect and $\Tilde{\Omega}_q = \Omega_q - i\gamma_q$ is the complex frequency of the mode in the $q^\text{th}$ cavity. We obtain these QMs and their complex frequencies via FDTD simulations. 
These modes are inherently lossy, with the effects of loss captured by the complex nature of the system QM frequencies: the real part, $\Omega_q$, gives the resonant frequency and the imaginary part, $\gamma_q$, describes the energy leakage out of the $q^\text{th}$ defect mode. The quality factor of the mode is given simply by $Q_q=\Omega_q/2\gamma_q$. 

\begin{figure}
    \centering
    \includegraphics[width=0.48\textwidth]{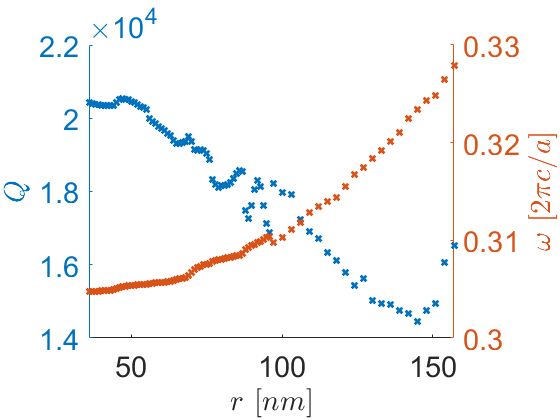}
    \caption{\justifying Single-defect QM frequency and quality factor as a function of defect air hole radius for the chosen PC. Computed with ANSYS Lumerical FDTD \cite{FDTD}.}
    \label{fig:freq_vs_r}
\end{figure}

In Fig. \ref{fig:freq_vs_r}, we plot the resonant frequency and quality factor of a single-defect cavity as a function of defect-hole radius. Note that by changing the hole radius from 50 nm to 150 nm, we can tune the resonant frequency by about 7\%, while the quality factor ranges from about 14,500 to about 20,500. This tunability allows us to design the RS, the CROWs, and the coupling cavities to achieve our overall design goals. 

The $\textbf{M}_q(\textbf{r})$ act as a basis for full system QMs that have frequencies well inside the chosen band gap of the PC. We use a tight-binding approach and  expand the system QMs, $\textbf{N}_\mu(\textbf{r})$, in terms these as \cite{Fussell2008}
\begin{align}\label{eq:TBexpansion}
    \textbf{N}_\mu(\textbf{r}) = \sum_q v_{q\mu}\textbf{M}_q(\textbf{r}),
\end{align}
where the $v_{q\mu}$ are the expansion coefficients for the $\mu^{th}$ system QM. The positive-frequency part of the electric field in the $\mu^\text{th}$ system QM then takes the form
\begin{align}
    \label{eq:systemQM_E+}
    \textbf{E}_\mu^+(\textbf{r},t) = \sqrt{\frac{\hbar\omega_\mu}{2\epsilon_0}}\textbf{N}_\mu(\textbf{r})e^{-i\Tilde{\omega}_\mu t},
\end{align}
where $\Tilde{\omega}_\mu=\omega_\mu - i\gamma_\mu$ is the complex frequency of the $\mu^\text{th}$ system QM. The $\textbf{N}_\mu(\textbf{r})$ satisfy the Helmholtz equation
\begin{align}\label{eq:helmholtz_cc}
    \curl\curl\textbf{N}_\mu(\textbf{r}) - \frac{\Tilde{\omega}_\mu^2}{c^2}\epsilon(\textbf{r})\textbf{N}_\mu(\textbf{r}) = 0,
\end{align}
where $\epsilon(\textbf{r})$ is the dielectric function of the full system. 

Eqs. (\ref{eq:helmholtz_iso}), (\ref{eq:TBexpansion}), and (\ref{eq:helmholtz_cc}) lead to the generalized eigenvalue problem \cite{Fussell2008}
\begin{align}
    \label{eq:TBEVP}
    \mat{A}\mat{\Omega}\mat{V} = \mat{\Lambda}\mat{B}\mat{V},
\end{align}
where $\mat{\Omega} \equiv \text{Diag}(\Tilde{\Omega}_q^2)$, $\mat{\Lambda} \equiv \text{Diag}(\Tilde{\omega}_\mu^2)$, $\mat{V}$ is a square matrix whose elements are the $v_{q\mu}$ from Eq. (\ref{eq:TBexpansion}), and $\mat{A}$ and $\mat{B}$ are respectively the overlap and coupling matrices for the coupled-cavity structure, whose elements are defined as:
\begin{align}
    A_{pq} &= \int d^3\textbf{r}\epsilon_p(\textbf{r})\textbf{M}_p^*(\textbf{r})\cdot\textbf{M}_q(\textbf{r}),\label{eq:AIntegral}\\
    B_{pq} &= \int d^3\textbf{r}\epsilon(\textbf{r})\textbf{M}_p^*(\textbf{r})\cdot\textbf{M}_q(\textbf{r}).\label{eq:BIntegral}
\end{align}
Note that all independent quantities in Eqs. (\ref{eq:TBEVP})-(\ref{eq:BIntegral}) can be obtained from FDTD/FEA simulations of individual defects. The number of required FDTD simulations is equal to the number of unique cavities, rather than the number of total cavities. In the designed system, there are 184 total cavities, but only 7 unique cavities.
We will use the full-system QMs $\textbf{N}_\mu(\textbf{r})$ as a basis for the classical time evolution of the pump field and the quantum evolution of the generated light.

The inner product between the system QMs is defined as
\begin{align}
    \bra{\textbf{N}_\mu}\ket{\textbf{N}_\nu} &\equiv \int_\infty d^3 \textbf{r}\epsilon(\textbf{r}) \textbf{N}_\mu^*(\textbf{r})\cdot\textbf{N}_\nu(\textbf{r})\\
    & = V_\mu^\dagger \mat{B} V_\nu\label{eq:QMinnerprod}\\
    & = O_{\mu\nu},\label{eq:OMatrixElems}
\end{align}
where $V_\mu$ is the $\mu^\text{th}$ column of $\mat{V}$. The $O_{\mu\nu}$ are elements of the square Hermitian matrix $\textbf{O}$, called the QM overlap matrix. We plot the overlaps between the full-system QMs in Fig. \ref{fig:FullSystemOverlaps}, and discuss their implications in Sec. \ref{sec:III}.

\subsubsection{Resonant Structure Quasi-Modes}
The first subsystem that we focus on is the RS. The RS QMs are computed in the same manner as those of the full structure, but we only include the cavities comprising the RS when calculating these QMs. 
The positive frequency part of light in the $j^\text{th}$ RS QM takes the form

\begin{align}
    \label{eq:pos_freq_E_RSQM}
    \textbf{E}_j^+(\textbf{r},t)=i\sqrt{\frac{\hbar\omega_j}{2\epsilon_0}}\textbf{C}_j(\textbf{r})e^{-i\comega_j t},
\end{align}
where $\textbf{C}_j(\textbf{r})$ is $j^{th}$ QM ($j=\{I,P,S\}$) of the RS and $\comega_j \equiv\omega_j-i\gamma_j$ is its complex frequency. 
The $\textbf{C}_j(\textbf{r})$ satisfy the Helmholtz equation 
\begin{align}
    \curl\curl\textbf{C}_j(\textbf{r})-\frac{\comega_j^2}{c^2}\epsilon_\text{RS}(\textbf{r)}\textbf{C}_j(\textbf{r})=0,
\end{align}
where $\epsilon_\text{RS}(\textbf{r})$ is the dielectric function of the slab with only the RS defects present. 

In this case the RS is constructed from three single-defect cavities arranged in a line (see Fig. \ref{fig:PCCCS}). 
Following Section \ref{sec:IIIA1}, the tight-binding eigenvalue problem for the RS QMs is
\begin{align}
    \label{eq:TBEVP_RS}
    \mat{A}_\text{RS}\mat{\Omega}_\text{RS}\mat{U}= \mat{\Lambda}_\text{RS}\mat{B}_\text{RS}\mat{U},
\end{align}
where $\mat{A}_\text{RS}$ and $\mat{B}_\text{RS}$ have elements respectively given by Eqs. (\ref{eq:AIntegral}) and (\ref{eq:BIntegral}) with $\epsilon \xrightarrow{}\epsilon_\text{RS}$ and the elements, $u_{qj}$ of $\mat{U}$ are the expansion coefficients of the RS QMs according to
\begin{align}
    \label{eq:Cexpansion}
    \textbf{C}_j(\textbf{r})=\sum_q u_{qj} \textbf{M}_q(\textbf{r}),
\end{align}
where the sum is only over the three defect modes in the RS.

\subsubsection{Bloch Modes of an Infinite CROW}\label{sec:NNTB}
The remaining subsystems are the three CROWs. To understand the modes of the finite CROWS in the system, it is useful to first determine the Bloch QMs of infinite CROWs. In the nearest-neighbor tight-binding (NNTB) approximation, these modes can be expanded in the basis of single-defect cavity QMs as \cite{Fussell2008}
\begin{align}
    \textbf{F}_k(\textbf{r})=\frac{1}{\sqrt{N}}\sum_q e^{ik q d}\textbf{M}_q(\textbf{r}),
\end{align}
where $d=3a$ is the CROW period and $N$ is the number of cavities in the CROW, which is taken to be of length $L=Nd$. In the limit that $N\rightarrow\infty$,
the solution to the generalized eigenvalue problem for the CROW in the NNTB gives \cite{Seifoory2019}
\begin{align}
    \tilde{\omega}_k=\tilde{\Omega}_0\left[1+(A_{01}-B_{01})\cos(kd)\right], 
\end{align}
for the complex frequency of the Bloch mode with Bloch wavevector $k$, where $\tilde{\Omega}_0$ is the complex frequency of all of the identical single cavity modes in the CROW and $A_{01}$ and $B_{01}$ are overlaps given in Eqs. (\ref{eq:AIntegral}) and (\ref{eq:BIntegral}) for neighboring single-cavity modes in the CROW structures. The radius of the defect holes in a given CROW is chosen to match the center frequency of that CROW band to the frequency of the corresponding RS QM ($P,S,I$). 
In Fig. \ref{fig:crowbands}, we plot the real part of the CROW frequencies as a function of the Bloch wavenumber, $k$, for all three CROWs.

The decay rate of light out of the resonant structure into the three CROWs is controlled by two \textit{coupling cavities} located between the resonant structure and the signal and idler CROWs (see Fig. \ref{fig:PCCCS}). The signal and idler coupling cavities have air hole radii of 55 and 91 nm respectively. As we shall discuss in more detail in Section \ref{sec:III}, these cavities tune the coupling between the resonant structure and the signal and idler CROWs. We have chosen the coupling cavities such that the loaded decay rates of the signal and idler RS QMs are similar. The objective is to strongly couple the signal and idler CROWs to the RS to allow entanglement to persist in the CROWs without compromising the nonlinear generation process. 

\section{Pump Evolution}\label{AppB}

In this appendix, we describe how we model the evolution of the pump light in the full system and calculate the coherent state amplitude, $\alpha_P(t)$, of the pump in the RS pump mode. The pump at time $t=0$  starts in a coherent (classical) state in the pump CROW.  We choose the initial state of the pump to be a Gaussian in the Bloch modes, centered at $k_0 =-\frac{\pi}{2d}$. With this choice, the propagation speed is maximized, there is no waveguide group velocity dispersion, and the frequency is centered at the pump frequency of the RS: $\omega_{k_0}=\Omega_0=\omega_P$.  
Thus, the positive frequency part of the electric field of the pump light at time $t=0$ is given by
\begin{align}
    \label{eq:Appendix_Epexp}
    \textbf{E}_P^{(+)}(\textbf{r},t=0)=i\sum_k\alpha_P^{(0)}(k)\sqrt{\frac{\hbar\tilde{\omega}_k}{2\epsilon_0}}\textbf{F}_k(\textbf{r}),
\end{align}
where the $\alpha_P^{(0)}(k)$ are the expansion coefficients, which take the form
\begin{align}
    \label{eq:app_alpha_P}
    \alpha^{(0)}_P(k)=\sqrt{\frac{\kappa d n_p}{\sqrt{2\pi}}} e^{-(k-k_0)^2/\kappa^2}e^{-iq_0 kd},
\end{align}
where $n_p$ is the total number of photons initially in the pump pulse, $q_0$ is the cavity on which the pulse is centered at $t=0$, and $\kappa$ is the $k$-space width parameter of the pulse. 

Using Eq. (\ref{eq:app_alpha_P}) in Eq. (\ref{eq:Appendix_Epexp}) and evaluating the sum over k in the limit that $N$ is very large, we obtain after some work  
\begin{align}
    \label{eq:E_t0_M_q}
    \textbf{E}_P^+(\textbf{r},0) =i \sqrt{\frac{\hbar\omega_P}{2\epsilon_0}} \sum_{q=1}^N x_q \textbf{M}_q(\textbf{r}),
\end{align}
where  
\begin{align}
    \label{eq:x_q}
    x_q = \sqrt{\frac{\kappa dn_p}{\sqrt{2\pi}}} e^{-\kappa^2d^2(q-q_0)^2/4}e^{i(q-q_0)k_0 d}.
\end{align}

Now, to determine the time evolution of the pump in the full system, we need to expand the pump field at $t=0$ in terms of the QMs of the full system, since we know that these evolve in time according to Eq. (\ref{eq:systemQM_E+}). We start by inverting Eq. (\ref{eq:TBexpansion}) to obtain the single cavity modes in terms of the full system modes:
\begin{align}
{\textbf{M}}_q(\textbf{r}) = \sum_\mu {w}_{\mu q} {\textbf{N}}_\mu(\textbf{r}),    
\end{align}
where $w_{\mu q}$ are the elements of $\mat{W} = \mat{V}^{-1}$. Using this in Eq. (\ref{eq:E_t0_M_q}) and using Eq. (\ref{eq:x_q}), we obtain
\begin{align}
    \label{eq:E_t_N_mu}
    \textbf{E}_P^+(\textbf{r},t) =i \sqrt{\frac{\hbar\omega_P}{2\epsilon_0}} \sum_{q=1}^N x_q \sum_\mu{w}_{\mu q}\textbf{N}_{\mu}(\textbf{r})e^{-i\tilde{\omega}_\mu t}.
\end{align}
Finally, using Eq. (\ref{eq:TBexpansion}) to express the full system modes once again in terms of the cavity modes, we obtain
\begin{align}
    \label{eq:E_t_M_q}
    \textbf{E}_P^+(\textbf{r},t) =i \sqrt{\frac{\hbar\omega_P}{2\epsilon_0}} \sum_{p=1}^N y_p(t) \textbf{M}_{p}(\textbf{r}),
\end{align}
where
\begin{align}
    \label{eq:y_p}
    y_p(t) \equiv \sum_q x_q \sum_\mu{w}_{\mu q}{v}_{p\mu}e^{-i\tilde{\omega}_\mu t}.
\end{align}
Eqs. (\ref{eq:x_q}), (\ref{eq:E_t_M_q}), and (\ref{eq:y_p}) may be used to reconstruct the electric field of the pump in the full structure as a function of time. 

The time-dependent coherent state amplitude, $\alpha_{P}(t)$, in the RS pump mode is obtained by projecting $\textbf{E}_P^+(\textbf{r},t)$ onto the pump QM of the RS to obtain 
\begin{align}
    \label{eq:Epdef}
    \alpha_P(t) &=  -i\sqrt{\frac{2\epsilon_0}{\hbar\omega_P}} \int d^3\textbf{r}\epsilon_{RS}(\textbf{r})\textbf{C}_{P}^*(\textbf{r})\cdot \textbf{E}_{P}^+(\textbf{r},t)\nonumber\\
    &\approx \sum_{q,q'} u^*_{qP}B_{qq'}y_{q'}(t).
\end{align}
We note that the second line is approximate since in the previous line we use $\epsilon_\text{RS}$, while $\mat{B}$ was computed using $\epsilon$. 

\section{Nonlinear Generation in the Resonant Structure}\label{AppC}
In this appendix, we present the coupled differential equations used to determine the state parameters of the STS in the RS. This follows the work of Vendromin and Dignam  \cite{Vendromin2021}. 
Writing the nonlinear driving term as
\begin{equation}
    \alpha_{P}^2(t)\chi_\text{eff}=|\alpha_{P}^2(t)\chi_\text{eff} |e^{-i\omega_P(t-t_0)}e^{i\beta (t)},
\end{equation}
we find, for our choice of initial pulse in the structure we have designed, that $\beta$ is independent of $t$ over the time range for which the pump is non-negligible.
When $\beta$ is independent of $t$ and the initial state is the vacuum state, it can be shown that the squeezing phase evolves simply according to \cite{Vendromin2021} 
\begin{align}
    \label{eq:squeezingPhase}
    \theta(\Tilde{t}) = -2\omega_P(\Tilde{t}-\Tilde{t}_0)-\frac{\pi}{2}+\beta,
\end{align}
and the squeezing amplitude and thermal photon numbers are governed by the following set of coupled first-order differential equations \cite{Vendromin2020}:
\begin{align}
    &\frac{dr}{d\Tilde{t}}  = \frac{g(\Tilde{t})}{2} - \frac{\sinh(2r)}{n^\text{th}_S + n^\text{th}_I + 1}\qty(1+\zeta(n^\text{th}_I-n^\text{th}_S)),\label{DynamicsDE1}\\
    &\frac{dn^\text{th}_S}{d\Tilde{t}}  = n^\text{th}_S((1-\zeta)\sinh^2r - (1+\zeta)\cosh^2r)\nonumber\\
    &\quad\quad+(1-\zeta)\sinh^2r,\label{DynamicsDE2}\\
    &\frac{dn^\text{th}_I}{d\Tilde{t}}  = n^\text{th}_I((1+\zeta)\sinh^2r - (1-\zeta)\cosh^2r)\nonumber\\
    &\quad\quad+(1+\zeta)\sinh^2r,\label{DynamicsDE3}    
\end{align}
where we employ the dimensionless time units $\Tilde{t} = \Gamma_+t$ with $ \Gamma{\pm} = (\Gamma_S \pm \Gamma_I)/2$. The parameter $\zeta = \Gamma_-/\Gamma_+$, quantifies the difference in the loss rates in the two modes and is the pump strength defined in Eq. (\ref{eq:pumpingStrength}).  Eqs. (\ref{DynamicsDE1}) to (\ref{DynamicsDE3}) are solved numerically using a fourth-order Runge-Kutta method, with the initial conditions: $r(0)=0$, $n^\text{th}_S(0)=0$, $n^\text{th}_I(0)=0$.

\section{Quantum State Evolution in the Output CROWs}\label{AppD}
In this Appendix, we model the free quantum evolution of the 2MSTS in the full system in the presence of loss but without the pump and compute the time-dependent expectation value of the CV between output CROW cavities. We accomplish this by first solving the adjoint master equation for the time-evolution of the full-system QM ladder operators, and then use the initial state of the system at time $\tilde{t}=\tilde{t}_0$ to calculate time dependence of the photon number and correlation variance. 

Consider the CV, $\Delta^2(p_S,\tilde{t}_S;p_I,\Tilde{t}_I)$ between two cavities, one each in the signal and idler CROW, denoted $p_S$ and $p_I$ respectively (see Fig. \ref{fig:PCCCS}), where we allow the measurement times $\Tilde{t}_{S}$ and $\Tilde{t}_{I}$ to be different. This CV may be expressed in terms of the quadrature operators in the two cavities as
\begin{align}
    &\Delta^2(p_S,\tilde{t}_S;p_I,\Tilde{t}_I) = \label{eq:outputCV}\\
    &\expval{\qty[X_{p_S}(\Tilde{t}_S)-X_{p_I}(\Tilde{t}_I)]^2}\nonumber\\
    &+\expval{\qty[Y_{p_S}(\Tilde{t}_S)+Y_{p_I}(\Tilde{t}_I)]^2},\nonumber
\end{align}
where
\begin{align}
    X_q(t) &= \frac{1}{2}\qty(a_q(t)  + a_q^\dagger(t)),\label{eq:X_q_t}\\
    Y_q(t) &= \frac{1}{2i}\qty(a_q(t) - a_q^\dagger(t)).\label{eq:Y_q_t}
\end{align}


Below, we shall write the time-dependent photon number and correlation variance in terms of the $a$-operators. First, though, we must determine the time-evolution of the $c$-operators, since, as we shall see, it is these (and not the $b$-operators) that display simple single-frequency time-dependence. We begin with the adjoint master equation, which describes the time-evolution of a Heisenberg picture operator $A_H$ in terms of the free Hamiltonian in Eq. (\ref{eq:H_free}) \cite{KamandarDezfouli2014}:
\begin{align}
    \label{eq:Adjoint}
    \dv{A_H}{\Tilde{t}}&=-\frac{i}{\hbar}A_H H_\text{free} + \frac{i}{\hbar}H_\text{free} ^{\dagger} A_H \nonumber\\&+i\sum_\mu\qty(\comega_\mu b_\mu^\dagger A_H c_\mu - \comega^*_\mu c_\mu^\dagger A_H b_\mu).
\end{align}

Taking $A_H\xrightarrow{}c_\nu$ we obtain, after some algebra,
\begin{align}
    \label{eq:Appendix_TE_c}
c_\nu(\Tilde{t}) = c_\nu e^{-i\Tilde{\omega}_\nu (\Tilde{t}-\tilde{t}_0)},
\end{align}
where $c_\nu\equiv c_\nu(\Tilde{t}_0)$, displaying the simple behavior we expected. For $c^\dagger(\Tilde{t})$ we simply take the Hermitian conjugate of Eq. (\ref{eq:Appendix_TE_c}). More generally, it is easy to show that for any normally-ordered set of $c-$operators, the evolution of each operator is simply given by the result shown in Eq. (\ref{eq:Appendix_TE_c}). The time-evolution of the $b_\nu$'s is found via construction from the $c_\nu$'s to be \cite{Dignam2012}
\begin{align}
    \label{eq:TE_b}
    &b_\mu(\Tilde{t})\equiv \sum_{\nu=1}^NO_{\nu\mu}c_\nu(\Tilde{t}),
\end{align}
where again $b^\dagger(\Tilde{t})$ is obtained via the Hermitian conjugate of Eq. (\ref{eq:TE_b}).
The time evolution of the single-cavity QM operators  (the $a$-operators), obtained using Eqs. (\ref{eq:single_cavity_inverse_tb}),  (\ref{eq:TE_c}), and (\ref{eq:TE_b}), is found to be
\begin{align}    
    \vec{a}(\Tilde{t})&=\mat{W}^\dagger\mat{O}\mat{F}(\Tilde{t})\mat{P}\mat{V}^\dagger \vec{a}\\
    &=\mat{\Phi}(\Tilde{t})\vec{a},\label{eq:Appendix_TE_a}
\end{align}
where $\vec{a}\equiv\qty(a_1,a_2,...,a_p)^T$ and $\mat{F}(\Tilde{t})=\text{diag}\qty(e^{-i\comega_1(\Tilde{t}-\tilde{t}_0)},e^{-i\comega_2(\Tilde{t}-\tilde{t}_0)}, ... ,e^{-i\comega_\nu (\Tilde{t}-\tilde{t}_0)})$. Inserting Eq. (\ref{eq:TE_a}) into Eqs. (\ref{eq:X_q_t}, \ref{eq:Y_q_t}) and substituting the result into Eq. (\ref{eq:outputCV}) yields, after some algebra
\begin{align}
    &\Delta^2(p_S,\Tilde{t}_S;p_I,\Tilde{t}_I) =\label{eq_a:outputCV_simplified}\\
    &\frac{1}{2}(B_{p_{S} p_{S}} + B_{p_{I} p_{I}})\nonumber\\
    &+\sum_{q,q'=1}^N [\phi_{p_{S}q}^*(\Tilde{t}_S)\phi_{p_{S}q'}(\Tilde{t}_S)\nonumber\\
    &+ \phi_{p_{I}q}^*(\Tilde{t}_I)\phi_{p_{I}q'}(\Tilde{t}_I)]\expval{a^\dagger_q a_{q'}}\nonumber\\
    &-2\Re{\sum_{q,q'=1}^N\phi_{p_{S}q}(\Tilde{t}_s)\phi_{p_{I}q'}(\Tilde{t}_I)\expval{a_q a_{q'}}}.\nonumber
\end{align}
In Eq. (\ref{eq_a:outputCV_simplified}), all the time-dependence is contained within the $\phi_{pq}(\tilde{t})$ elements, and the only quantities still unknown are the $\expval{a^\dagger_p a_q}$ and $\expval{a_p a_q}$ ensemble averages, which we shall now compute. These ensemble averages are evaluated at time $\tilde{t}=\tilde{t}_0$ (see Fig. \ref{fig:gp}). We assume that at this time, the only light in the system is the generated light which exists exclusively in the signal and idler QMs of the RS. This enables us to write: 
\begin{align}
    \expval{a_p a_q} & = \sum_{j,k}{\sigma}^*_{pj} {\sigma}^*_{qk}\expval{d_j d_{k}},\label{eq:aa_sigmanotation}\\
    \expval{a_p^\dagger a_q} & = \sum_{j,k}\sigma_{pj} {\sigma}^*_{qk}\expval{d_j^\dagger d_{k}},\label{eq:adaga_sigmanotation}
\end{align}
where $p$ and $q$ are now cavities in the RS, the $d_j$ ($d_j^\dagger$) are the annihilation (creation) operators for the $j^\text{th}$ RS QM, and the $\sigma_{pj}$ are the matrix elements of the matrix $\mat{\Sigma}=\mat{U}^{-1}$ (see Eq. (\ref{eq:TBEVP_RS}). In Eqs. (\ref{eq:aa_sigmanotation}-\ref{eq:adaga_sigmanotation}), $j$ and $k$ run through only the signal ($j=S$) and idler ($k = I$) modes of the RS.

We now directly evaluate the ensemble averages in Eqs. (\ref{eq:aa_sigmanotation}, \ref{eq:adaga_sigmanotation}). Using the density operator from Eq. (\ref{eq:2MSTS_densityOp}), the ensemble average for a general operator $A$, acting on a two-mode squeezed state can be expressed as:
\begin{align}
    \expval{A} & = \Tr{A\rho}\nonumber\\
    &= \Tr{AS_2(\xi)\rho^\text{th} S^\dagger_2(\xi)}\nonumber\\
    &= \Tr{S_2^\dagger(\xi)A S_2(\xi) \rho^\text{th}}\label{appD_error1}. 
\end{align}
Using the standard relations \cite{GerryKnight}, we obtain:
\begin{align}
    S_2^\dagger(\xi) d_S S_2(\xi) & = d_S\cosh r - d_I^\dagger e^{i\theta}\sinh r, \label{appD_error2}\\
    S_2^\dagger(\xi) d_I S_2(\xi) & = d_I\cosh r - d_S^\dagger e^{i\theta}\sinh r \label{eq:Sc2S},\\
    S_2^\dagger(\xi) d_S^\dagger S_2(\xi) & = d_S^\dagger\cosh r - d_I e^{-i\theta}\sinh r, \label{eq:Sc3S}\\
    S_2^\dagger(\xi) d_I^\dagger S_2(\xi) & = d_I^\dagger\cosh r - d_S e^{-i\theta}\sinh r \label{appD_error3}.
\end{align}

Thus, from Eq. (\ref{appD_error1}) and Eqs. (\ref{appD_error2} - \ref{appD_error3}), we can evaluate Eqs. (\ref{eq:aa_sigmanotation}, \ref{eq:adaga_sigmanotation}). After some algebra, we obtain:
\begin{align}
    &\expval{a_p a_q} =\label{appD_error4}\\
    &-\qty({\sigma}^*_{pS} {\sigma}^*_{qI}+ {\sigma}^*_{pI} {\sigma}^*_{qS})\nonumber\\
    &\times \qty(n^{th}_S+n^{th}_I+1)e^{i\theta}\cosh r, \sinh r,\nonumber
\end{align}
and
\begin{align}
    &\expval{a_p^\dagger a_q} =\label{appD_error5}\\
    &\sigma_{pS} {\sigma}^*_{qS} \qty(n^{th}_S\cosh^2r + (1+n^{th}_I)\sinh^2r) \nonumber\\
    &+ \sigma_{pI} {\sigma}^*_{qI} \qty(n^{th}_I\cosh^2r + (1+n^{th}_S)\sinh^2 r),\nonumber
\end{align}
where $r$, $n^\text{th}_S$, and $n^\text{th}_I$ are evaluated at $\tilde{t}=\tilde{t}_0$. Finally, the combination of Eqs. (\ref{appD_error4}, \ref{appD_error5}) with Eq. (\ref{eq_a:outputCV_simplified}) yields an expression for the correlation variance between two cavities in the signal and idler output CROWs. The preceding analysis also yields an expression for the average photon number in the output CROW cavities. Since Eq. (\ref{eq:adaga}) is normally ordered, we may simply use Eq. (\ref{eq:Appendix_TE_a}) to obtain Eq. (\ref{eq:TE_PN}).

\end{document}